\documentclass[a4paper,11pt]{article}
\pdfoutput=1
\usepackage{jheppub}
\usepackage[toc,page]{appendix}
\usepackage{graphicx}
\usepackage{amsmath,amssymb,mathrsfs}
\usepackage{bbm}
\usepackage{color}
\usepackage{xcolor}
\usepackage{dsfont} 
\usepackage{cancel}
\usepackage{dsfont}
\usepackage{epstopdf}
\usepackage{epsfig}
\usepackage{bm}
\usepackage{dcolumn}
\usepackage{hyperref}
\usepackage{enumitem}
\usepackage{multirow}
\usepackage{lineno}
\usepackage{mathtools,slashed}
\usepackage{url}
\usepackage{lineno}
\usepackage{wrapfig}
\usepackage{subfigure}

\usepackage{tikz-feynman}
\tikzset{
photon/.style={decorate, decoration={snake,amplitude=2pt, segment length=5pt}, draw=black},
particle/.style={draw=black, postaction={decorate}, decoration={markings,mark=at position .5 with {\arrow[draw=black]{>}}}},
antiparticle/.style={draw=black, postaction={decorate}, decoration={markings,mark=at position .5 with {\arrow[draw=black]{>}}}},
gluon/.style={decorate, draw=black, decoration={coil,amplitude=4pt, segment length=5pt}}
goldstone/.style={draw=green,postaction={decorate},decoration={markings,mark=at position .5 with {\arrow[draw=blue]{>}}}}
}

\newcommand{\ben}{\begin{enumerate}}
\newcommand{\een}{\end{enumerate}}
\newcommand{\bit}{\begin{itemize}}
\newcommand{\eit}{\end{itemize}}

\newcommand{\beqa}{\begin{eqnarray}}
\newcommand{\eeqa}{\end{eqnarray}}
\newcommand{\beq}{\begin{equation}}
\newcommand{\eeq}{\end{equation}}
\newcommand{\bay}{\begin{array}}
\newcommand{\eay}{\end{array}}

\def\ifmath#1{\relax\ifmmode #1\else $#1$\fi}

\def\gsim{\ \rlap{\raise 3pt \hbox{$>$}}{\lower 3pt \hbox{$\sim$}}\ }
\def\lsim{\ \rlap{\raise 3pt \hbox{$<$}}{\lower 3pt \hbox{$\sim$}}\ }

\def\ls#1{\ifmath{_{\lower1.5pt\hbox{$\scriptstyle #1$}}}}
\def\lsup#1{^{\lower 6pt\hbox{$\scriptstyle#1$}}}

\def\Tr{{{\rm Tr}}}

\def\bracket#1#2 {\mathinner{\langle{#1}|{#2}\rangle}}

\def\bracket#1#2 {\mathinner{\langle{#1}|{#2}\rangle}}

\newcommand{\be}{\begin{equation}}
\newcommand{\ee}{\end{equation}}
\newcommand{\bea}{\begin{eqnarray}}
\newcommand{\eea}{\end{eqnarray}}

\newcommand{\red}[1]{{\color{red} #1}}

\graphicspath{{figs/}}

\begin{document}

\title{The Scalar Singlet Extension of the Standard Model: Gravitational Waves versus Baryogenesis}

\abstract{  
We study the possible gravitational wave signal and the viability of baryogenesis arising from the electroweak phase transition in an extension of the Standard Model (SM) by a scalar singlet field without a $\mathbb{Z}_2$ symmetry. We first analyze the velocity of the expanding true-vacuum bubbles during the phase transition, confirming our previous finding in the unbroken $\mathbb{Z}_2$ symmetry scenario, where the bubble wall velocity can be computed from first principles only for weak transitions with strength parameters $\alpha \lesssim 0.05$, and the Chapman-Jouguet velocity defines the maximum velocity for which the wall is stopped by the friction from the plasma. We further provide an analytical approximation to the wall velocity in the general scalar singlet scenario without $\mathbb{Z}_2$ symmetry and test it against the results of a detailed calculation, finding good agreement. We show that in the singlet scenario with a spontaneously broken $\mathbb{Z}_2$ symmetry, the phase transition is always weak and we see no hope for baryogenesis. In contrast, in the case with explicit $\mathbb{Z}_2$ breaking there is a region of the parameter space producing a promising baryon yield in the presence of CP violating interactions via an effective operator involving the singlet scalar and the SM top quarks. Yet, we find that this region yields unobservable gravitational waves. Finally, we show that the promising region for baryogenesis in this model may be fully tested by direct searches for singlet-like scalars in di-boson final states at the HL-LHC, combined with present and future measurements of the electron electric dipole moment.
%
~~\\
~~\\
KCL-PH-TH/2022-48, CERN-TH-2022-150, IFT–UAM/CSIC–22-133

}

\author[1,2,3]{John~Ellis,}
\author[4]{Marek~Lewicki,}
\author[4]{Marco~Merchand,}
\author[5,6]{Jos\'e~Miguel~No,}
\author[4]{Mateusz~Zych}
\affiliation[1]{Department of Physics, King's College London, Strand, London WC2R 2LS, UK}
\affiliation[2]{Theoretical Physics Department, CERN, Geneva, Switzerland}
\affiliation[3]{National Institute of Chemical Physics \& Biophysics, R\"avala 10, 10143 Tallinn, Estonia}
\affiliation[4]{Faculty of Physics, University of Warsaw, ul.\ Pasteura 5, 02-093 Warsaw, Poland}
\affiliation[5]{Instituto de F\'isica Te\'orica, IFT-UAM/CSIC,
Cantoblanco, 28049 Madrid, Spain}
\affiliation[6]{Departamento de F\'isica Te\'orica, Universidad Aut\'onoma de Madrid, 28049 Madrid, Spain}
\emailAdd{John.Ellis@cern.ch}
\emailAdd{marek.lewicki@fuw.edu.pl}
\emailAdd{mmerchand@fuw.edu.pl}
\emailAdd{Josemiguel.no@uam.es}
\emailAdd{mateusz.zych@fuw.edu.pl}
\maketitle
\section{Introduction}

The lack of an explanation within the Standard Model (SM) for the observed baryon
asymmetry of the Universe is one of the key motivations for studying scenarios
for physics beyond the SM (BSM). One attractive scenario is that baryogenesis be
associated with the electroweak scale, potentially maximizing its testability
in laboratory experiments~\cite{Kuzmin:1985mm,Cohen:1993nk,Rubakov:1996vz, Morrissey:2012db}. However, electroweak baryogenesis (EWBG) would
require a first-order phase transition (FOPT) at the electroweak scale, which
does not occur in the SM~\cite{Kajantie:1996mn}. If it were to occur in some BSM scenario, collisions between bubbles of the low-energy vacuum and the ensuing turbulence and sound waves in the primordial
plasma might have generated a stochastic cosmological background of gravitational
waves (GWs) large enough to be detectable in future experiments such as LISA~\cite{Caprini:2015zlo,Caprini:2019egz} or AEDGE~\cite{AEDGE:2019nxb,Badurina:2021rgt}. 

This possibility has stimulated increased interest in BSM scenarios involving a FOPT~\cite{Dorsch:2014qpa,Jaeckel:2016jlh,Jinno:2016knw,Chala:2016ykx,Chala:2018opy,Artymowski:2016tme,Hashino:2016xoj,Vaskonen:2016yiu,Dorsch:2016nrg,Beniwal:2017eik,Baldes:2017rcu,Marzola:2017jzl,Kang:2017mkl,Iso:2017uuu,Chala:2018ari,Bruggisser:2018mrt,Megias:2018sxv,Croon:2018erz,Alves:2018jsw,Baratella:2018pxi,Angelescu:2018dkk,Croon:2018kqn,Brdar:2018num,Beniwal:2018hyi,Breitbach:2018ddu,Marzo:2018nov,Baldes:2018emh,Prokopec:2018tnq,Fairbairn:2019xog,Helmboldt:2019pan,Dev:2019njv,Ellis:2019flb,Jinno:2019bxw,Ellis:2019tjf,Azatov:2019png,vonHarling:2019gme,DelleRose:2019pgi,Mancha:2020fzw,Vanvlasselaer:2020niz,Giese:2020znk,Hoeche:2020rsg,Croon:2020cgk,Ares:2020lbt,Cai:2020djd,Bigazzi:2020avc,Wang:2020zlf,Baldes:2021aph,Yang:2022ghc,Zhou:2022mlz,Brzeminski:2022haa,Azatov:2022tii}.
Scenarios for electroweak baryogenesis exploit the fact that electroweak sphalerons are active and postulate
a CP-violating force that generates a matter-antimatter asymmetry in chemical potentials.
At the end of the transition when the sphaleron rate is suppressed, this
asymmetry freezes out into the net baryon number observed today.  
  
One of the simplest scenarios capable of producing a FOPT invokes a gauge singlet scalar in
addition to the SM~\cite{Espinosa:1993bs,Profumo:2007wc,Espinosa:2011ax}. In spite of its minimality, 
this scenario offers a vast array of possibilities for phenomenological studies,
including collider searches, sensitivities and constraints~\cite{Profumo:2007wc,Noble:2007kk,No:2013wsa,Curtin:2014jma,Huang:2017jws}, 
dark matter~\cite{Barger:2007im, He:2009yd, Gonderinger:2009jp,Cline:2013gha}, 
gravitational wave production~\cite{Croon:2018new} and baryogenesis~\cite{Espinosa:2011eu,Cline:2012hg}. 
In particular, the complementarity between collider searches and GW experiments for probing the 
parameter space of the model has been studied extensively~\cite{Alves:2018oct, Alves:2018jsw, Alves:2019igs,Alves:2020bpi}. 

Implementing EWBG within this scalar singlet extension of the SM requires the introduction of extra sources of CP violation. 
This is usually achieved by adding higher-dimension operators to the Lagrangian in order to keep the particle content minimal. 
However, these operators are subject to tight constraints from experimental upper bounds on electric dipole moments (EDMs),
unless they are made to vanish at zero temperature. To the best of our knowledge, the EWBG
computation has not been performed in the most generic scenario with a non-vanishing vev for the scalar singlet,
and computations of the baryon asymmetry have been focused
on the case in which the singlet has a vanishing vev.
On the other hand, studies of the potential synergy between colliders and GW experiments
for probing the region of parameter space compatible with EWBG have mainly focused 
on the scenario with non-vanishing singlet vev \cite{Alves:2018oct, Alves:2019igs, Alves:2020bpi}, while remaining agnostic about the source(s) of CP violation.~\footnote{See,
however,~\cite{Curtin:2014jma} for an exception.} 

The aim of the present paper is to plug this gap by investigating the viability of EWBG in the scalar singlet extension  of the SM using the most general parametrization of the scalar potential and assuming a non-vanishing vev for the singlet. 
As CP-violating source we include a dimension-five operator coupling the singlet field with the Higgs-top Yukawa interaction.  The final baryon yield is computed via the WKB formalism \cite{Cline:2000kb,Joyce:1994zt, Joyce:1994fu}.~\footnote{An alternative method is given by the vev-insertion approximation (VIA) \cite{Riotto:1995hh,Riotto:1997vy} which generically yields $\mathcal{O}(10) $ larger values for the baryon asymmetry although it was recently claimed to be not fully consistent~\cite{Kainulainen:2021oqs,Postma:2022dbr}. See Refs. \cite{Cline:2021dkf, Cline:2020jre} for comparative studies of these formalisms.} We find that, due to the presence of the higher dimensional operator involving the singlet and the SM top quarks (as needed here for baryogenesis), future LHC searches for singlet-like scalars in di-boson final states will be able to probe this baryogenesis scenario even in the limit of very small singlet-Higgs mixing.

We provide at the same time updated predictions for the GW signals taking into account a full computation of the properties of the bubble wall using the fluid equations of the {new formalism} of \cite{Laurent:2020gpg, Cline:2021iff}. Thus we extend our previous work~\cite{Lewicki:2021pgr} on the properties of bubble walls in SM-like thermal plasmas, showing that the same qualitative conclusions apply for this model. 

The rest of the paper is outlined as follows: In Sec.~\ref{sec:Veff} we introduce the model focusing on the effective potential. Sec.~\ref{sec:Phase Transition Dynamics} dynamics of the phase transition. In Sec.~\ref{sec:spontaneous_Z_2_breaking} we focus on the case of spontaneous $\mathbb{Z}_2$ symmetry breaking. We show that the transitions in that case are always too weak to give hope for baryogenesis, and move on to the case of explicit $\mathbb{Z}_2$ symmetry breaking in the remainder of the paper. 
Sec~\ref{sec:wall_properties} discusses the computation of the bubble wall properties as well as the analytical approximation of the bubble wall velocity and its accuracy.   
Using these results in Sec.~\ref{sec:GWs} we discuss the GW signals the model can produce. In Sec.~\ref{sec:Baryogenesis} we discuss our scenario for
baryogenesis, first setting out the 
phenomenological constraints on the CP-violating model and then discussing the possible magnitude of the cosmological baryon
asymmetry. We highlight the part of parameter space that yields results consistent with cosmology, and discuss the possible LHC and EDM probes of this region.
Sec.~\ref{Sec:conx} summarizes our conclusions.
The issues of vacuum stability and perturbativity are discussed in an Appendix.

\section{The Effective Potential of the Singlet Scalar Extension of the SM}\label{sec:Veff}

\subsection{Tree-level potential}

The model we consider in this paper has been studied previously in Refs.~\cite{Alves:2020bpi,Beniwal:2018hyi, Profumo:2007wc},
and the complementarity between its signatures in GW and collider experiments was studied in~\cite{Alves:2018jsw}.
The tree-level scalar potential of the model is
\begin{equation}
V_0(H,s) = -\mu_h^2H^{\dagger}H + \lambda_h (H^{\dagger}H)^2-\frac{1}{2}\mu_s^2 s^2 +\frac{1}{4}\lambda_s s^4+ \frac{1}{2}\lambda_{hs}H^{\dagger}H s^2 + \mu_{hs}H^{\dagger}Hs - \frac{1}{3}\mu_3 s^3, 
\label{potential}
\end{equation}
where  $H = (G^+ , \frac{h + i G^0}{\sqrt{2}})^\text{T}$ is the Higgs doublet with the SM vacuum expectation value (vev) $v=246.2$ GeV, 
while $s$ is the additional real scalar singlet. Eq.~\eqref{potential} is the most general formulation of the model, 
in which the $\mathbb{Z}_2$ symmetry of the potential corresponding to changing the sign of $s$ is not assumed. 
In a unitary gauge the potential (\ref{potential}) can be written as
\begin{equation}
V_0(h,s) = -\frac{1}{2}\mu_h^2 h^2 + \frac{1}{4}\lambda_h h^4 - \frac{1}{2}\mu_s^2 s^2 + \frac{1}{4}\lambda_{hs}h^2 s^2 + \frac{1}{4}\lambda_s s^4 + \frac{1}{2}\mu_{hs}h^2s - \frac{1}{3}\mu_3 s^3.
\end{equation}
The electroweak symmetry-breaking (EWSB) vacuum is located at the field values $(h,s) = (v,u)$, where
\begin{equation}
\begin{split}
    \frac{\partial V_0}{\partial h}\Big{|}_{(v,u)} &= -\mu_h^2 v + \lambda_h v^3 + \frac{1}{2}\lambda_{hs} vu^2 + \mu_{hs} vu = 0 \, ,\\
    \frac{\partial V_0}{\partial s}\Big{|}_{(v,u)} &= -\mu_s^2 u + \lambda_s u^3 + \frac{1}{2}\lambda_{hs} v^2u + \frac{1}{2}\mu_{hs}{v^2} - \mu_3 u^2 = 0 \, ,
\end{split}
\end{equation}
which lead to the following conditions
\begin{equation}
\begin{split}
    \mu^2_h &= \lambda_h v^2 +\frac{1}{2}\lambda_{hs} u^2 + \mu_{hs}u \, ,\\
    \mu^2_s &= \lambda_s u^2 + \frac{1}{2}\lambda_{hs} v^2 +\frac{1}{2}\mu_{hs}\frac{v^2}{u} - \mu_3 u \, .
\end{split}
\end{equation}
The elements of the scalar mass matrix $M$ are
\begin{equation}
\label{M}
\begin{split}
    M_{hh}^2 = \frac{\partial^2 V_0}{\partial h^2}\Big{|}_{(v,u)} &= -\mu_h^2 + 3\lambda_h v^2 +\frac{1}{2}\lambda_{hs}u^2 + \mu_{hs}u = \\
    &= 2\lambda_h v^2 \, ,\\
    M_{ss}^2 = \frac{\partial^2 V_0}{\partial s^2}\Big{|}_{(v,u)} &= -\mu_s^2 + 3\lambda_s u^2 +\frac{1}{2}\lambda_{hs}v^2 - 2\mu_{3}u=\\
    &= 2\lambda_s u^2 - \mu_{3}u - \frac{1}{2u}\mu_{hs}v^2 \, ,\\
    M_{hs}^2 = \frac{\partial^2 V_0}{\partial h \partial s}\Big{|}_{(v,u)} &= \lambda_{hs} vu + \mu_{hs} v \, .
    \end{split}
\end{equation}
In order to obtain the physical scalar masses, one diagonalizes the mass matrix and
introduces mass eigenstates $\varphi_1, \varphi_2$ given by
\begin{equation}
    \begin{pmatrix}
    \varphi_1\\
    \varphi_2\\
    \end{pmatrix}
    =
    \begin{pmatrix}
    \cos\theta & -\sin\theta\\
    \sin\theta & \cos\theta\\
    \end{pmatrix}
    \begin{pmatrix}
    h\\
    s\\
    \end{pmatrix} \, ,
\end{equation}
which satisfy 
\begin{equation}
\begin{split}
    &
    \begin{pmatrix}
    \varphi_1,\varphi_2\\
    \end{pmatrix}
    \begin{pmatrix}
    m^2_h & 0\\
    0 & m^2_s\\
    \end{pmatrix}
    \begin{pmatrix}
    \varphi_1\\
    \varphi_2\\
    \end{pmatrix}
    =
    \begin{pmatrix}
    h,s\\
    \end{pmatrix}
    \begin{pmatrix}
    M^2_{hh} & M^2_{hs}\\
    M^2_{sh} & M^2_{ss}\\
    \end{pmatrix}
    \begin{pmatrix}
    h\\
    s\\
    \end{pmatrix}
    =\\
    =&
    \begin{pmatrix}
    \varphi_1,\varphi_2\\
    \end{pmatrix}
    \begin{pmatrix}
    \cos\theta & -\sin\theta\\
    \sin\theta & \cos\theta\\
    \end{pmatrix}
    \begin{pmatrix}
    M^2_{hh} & M^2_{hs}\\
    M^2_{sh} & M^2_{ss}\\
    \end{pmatrix}
    \begin{pmatrix}
    \cos\theta & \sin\theta\\
    -\sin\theta & \cos\theta\\
    \end{pmatrix}
    \begin{pmatrix}
    \varphi_1\\
    \varphi_2\\
    \end{pmatrix} \, .
\end{split} 
\end{equation}
Comparing corresponding elements in these matrix equations, one obtains 
\begin{equation}
\begin{split}
m^2_h &= M^2_{hh}\cos^2\theta +M^2_{ss}\sin^2\theta - M^2_{hs}\sin2\theta \, , \\
m^2_s &= M^2_{hh}\sin^2\theta +M^2_{ss}\cos^2\theta + M^2_{hs}\sin2\theta \, , \\
0 &= -\frac{1}{2}(M_{ss}^2-M_{hh}^2)\sin 2\theta +M^2_{hs}\cos 2\theta \, .
\end{split}
\end{equation}
Inverting the system, we find
\begin{equation}
\begin{split}
M^2_{hh} & = m_h^2\cos^2\theta + m_s^2\sin^2\theta \, , \\
M^2_{ss} & = m_h^2\sin^2\theta + m_s^2\cos^2\theta \, , \\
M^2_{hs} &= - m_h^2\sin\theta\cos\theta + m_s^2\sin\theta\cos\theta \, .\\
\end{split}
\end{equation}
After simplifications and using (\ref{M}), we find the following conditions:
\begin{equation}
\begin{split}
\lambda_h &= \frac{1}{2 v^2}(m_h^2\cos^2\theta+m_s^2\sin^2\theta) \, , \\
\lambda_s &= \frac{1}{2u^2}(m_h^2\sin^2\theta + m_s^2\cos^2\theta + \mu_3 u +\frac{1}{2}\mu_{hs}\frac{v^2}{u}) \, , \\
\lambda_{hs} &= \frac{1}{vu}((m_s^2-m_h^2)\sin\theta\cos\theta - \mu_{hs}v) \, .
\end{split}
\end{equation}
Positivity imposes the following requirement in addition to $\lambda_h,\lambda_s>0$:
\begin{equation}
    \lambda_{hs}> -2\sqrt{\lambda_h \lambda_s}, \quad \text{if} \quad \lambda_{hs}<0 \, .
\end{equation}
In our numerical study of the parameter space we impose perturbativity of the quartic couplings by requiring they satisfy the conditions
\begin{equation}
    \lambda_{h}, \lambda_{s},  |\lambda_{hs}| \le 4\pi \, .
\end{equation}

\subsection{One-Loop Effective potential}

The full one-loop effective potential at finite temperature
can be represented in general as
\begin{equation}
V_{\textrm{eff}}(h,s,T)=V_0(h,s)+V_{\text{CW}}(h,s) + V_{\text{T}}(h,s,T) \, ,
\end{equation}
where $V_0$ is the tree-level potential, $V_{CW}$ denotes the 
one-loop corrections known as the Coleman-Weinberg potential~\cite{Weinberg:1973am} 
and $V_T$ represents the finite-temperature contribution.
Using the cutoff regularization and on-shell renormalization scheme,
the Coleman-Weinberg part is given by
\begin{equation}
V_{\text{CW}}(h,s) = \sum_i (-1)^{F_i}  \frac{d_i}{64 \pi^2} \left[ m_{i}^4(h,s) \left(  \log{\frac{m_{i}^2(h,s)}{m_{0i}^2}}  - \frac{3}{2}   \right) +2m_i^2(h,s) m_{0i}^2 \right] \, ,
\end{equation}
where the index $i$ runs over all particles contributing to the potential 
with $F_i=0$ ($1$) for bosons (fermions), $d_i$ is the number of 
degrees of freedom of the particle species, $m_i(h,s)$  is the 
field-dependent mass of particle $i$ and $m_{0i}$ its value in
the EW vacuum of the SM. The field-dependent masses are given 
in the SM by
\begin{equation}
m_W^2  =\frac{g^2}{4}h^2, \quad m_Z^2=\frac{g^2+g'^2}{4}h^2, \quad  m_t^2=\frac{y_t^2}{2}h^2 \, .  \label{field_dependent_masses}
\end{equation}
The thermally-corrected masses of the scalars are the eigenvalues of 
the thermally-corrected Hessian matrix
\begin{equation}
M^2 \rightarrow M^2 +
    \begin{pmatrix}
    \Pi_h(T) & 0\\
    0 &  \Pi_s(T)\\
    \end{pmatrix} \, ,
\end{equation}
while for the  longitudinal polarization states of vector bosons
\begin{equation}
    m_W^2 \rightarrow m_W^2 + \Pi_{W}(T)
\end{equation}
and 
\begin{equation}
M_{Z/\gamma}^2 =
    \begin{pmatrix}
    \frac{1}{4}g^2 h^2 + \frac{11}{6}g^2 T^2 & -\frac{1}{4}g g' h^2\\
    -\frac{1}{4}g g' h^2 & \frac{1}{4}g'^2 h^2 + \frac{11}{6}g'^2 T^2\\
    \end{pmatrix} \, .
\end{equation}
The thermal masses in this model are found to be \cite{Weinberg:1974hy}
\begin{equation}
 \Pi_h = \left( \frac{3g^2}{16} + \frac{g'^2}{16} + \frac{\lambda}{2} + \frac{y_t^2}{4} + \frac{\lambda_{hs}}{24}  \right) T^2 \, ,
\end{equation}
\begin{equation}
\Pi_s =\left(   \frac{\lambda_{hs}}{6} + \frac{\lambda_s}{4} \right)T^2 \, ,
\end{equation}
\begin{equation}
\Pi_{W} = \frac{11}{6}g^2 T^2 \, .
\end{equation}
These results are obtained by daisy resummation, which yields the following full
finite-temperature expression:
\begin{equation}
V_T(\phi, T) = \frac{T^4}{2\pi^2}  \sum_i d_i J_{\mp} \left( \frac{m_i(\phi)}{T} \right) \, ,
\end{equation}
where the $J_{\mp}$ functions are defined as  
\begin{equation}
J_{\mp}(x)  = \pm \int^{\infty}_{0} dy y^2 \log\left( 1 \mp e^{-\sqrt{y^2 + x^2}} \right) \, ,
\end{equation}
 the upper (lower) sign is for bosons (fermions) and $m_i$ refers to masses including the thermal corrections discussed above.

\section{Phase Transition Dynamics: Nucleation and Percolation} \label{sec:Phase Transition Dynamics}

The methods for computing the thermodynamic properties of a thermal 
FOPT in perturbative models are well established. 
One begins with the critical temperature $T_c$, defined to be the temperature 
at which multiple minima of the thermal effective potential become degenerate. 
One needs then to find the time of nucleation at which the 
probability of a true vacuum bubble forming within a horizon radius 
becomes significant, i.e.,
\be
N(T_n) = \int_{T_n}^{T_c} \frac{dT}{T} \frac{\Gamma(T)}{H(T)^4} = 1 \, ,
\ee
where 
\be
\Gamma(T) =\left(  \frac{S_3}{2\pi T}\right)^{3/2} T^4 e^{-S_3/T}
\ee
is the nucleation probability per unit time and volume, 
$S_3$ denotes the Euclidean action corresponding to the bounce solution 
and $H(T)$ is the Hubble expansion rate.

While computing the bounce for a model with a single scalar field is 
made relatively easy by use of a shooting algorithm, 
the task becomes significantly more onerous as the number of scalars increases. 
Moreover, one generally wishes to survey the full parameter space of the theory,
compounding the problem. In this paper we use the publicly available code {\tt cosmoTransitions}~\cite{Wainwright:2011kj}, which in principle can deal with an  arbitrary number of scalars. 

To leading-order accuracy and for temperatures close to the electroweak scale, 
the nucleation temperature can be obtained from the requirement
\be
\frac{S_3}{T_n} \approx 140 \, ,
\ee
which provides a good approximation for sufficiently weak transitions. 
This condition is already embedded in the public version of the 
{\tt cosmoTransitions} code, and we use this version for a 
preliminary survey of the parameter space in which a FOPT occurs. 
However, when the transition becomes too strong, the formula presented above 
is only indicative and a more careful treatment is necessary in order 
to evaluate if nucleation actually occurs. 

To assess if nucleation is possible we look for the solution to 
\be
\Gamma(T_n) = H_{\text{total}}(T_n)^4 \, ,
\ee
with the highest temperature, and we use the total Hubble rate
including the vacuum contribution, namely
\be
H^2_{\text{total}}(T) =  \frac{g_*(T) T^4}{90 \pi^2 M_{\text{Pl}}^2} + \frac{\Delta V(T)}{3 M_{\text{Pl}}^2},
\ee
where $g_*$ is the number of relativistic degrees of 
freedom~\cite{Saikawa:2018rcs} and $M_{\text{Pl}} = 2.4 \times 10^{18}$ GeV
is the reduced Planck mass. 

In addition to determining when nucleation is possible, an essential
question is whether the transition completes~\cite{Ellis:2018mja}. 
This can be answered by computing the temperature at which  the 
probability of a point remaining in the false vacuum drops below  0.71
and then verifying that the false vacuum volume 
is indeed shrinking at that temperature, i.e. 
\be
I(T) = \frac{4 \pi}{3} \int_{T}^{T_c} \frac{dT'}{H(T')} \Gamma(T') \frac{r(T,T')^3}{T'^4} = 0.34, \quad T \frac{dI(T)}{dT}<-3 \, , 
\ee 
where 
\be
\label{comoving_radius}
r(t,t') = v_w \int_{t'}^{t}  \frac{ d\tilde{t}}{a(\tilde{t})}
\ee
is the comoving radius of the bubble. The temperature at which the above 
two conditions are satisfied is referred to as the percolation temperature $T_p$.

\begin{figure}[!h]
\centering
\subfigure{
\includegraphics[width=7.2cm]{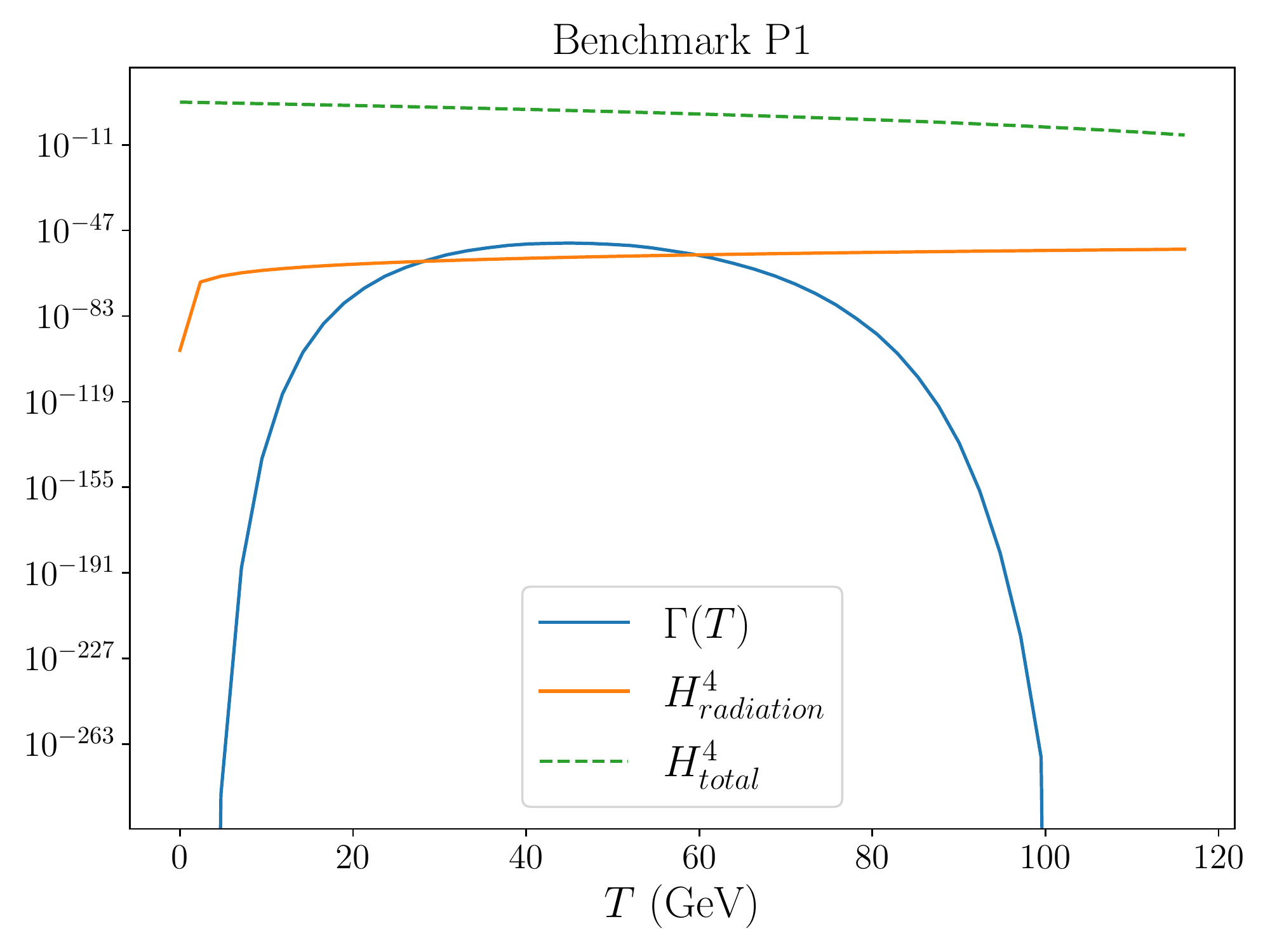}
}
\subfigure{
\includegraphics[width=7.2cm]{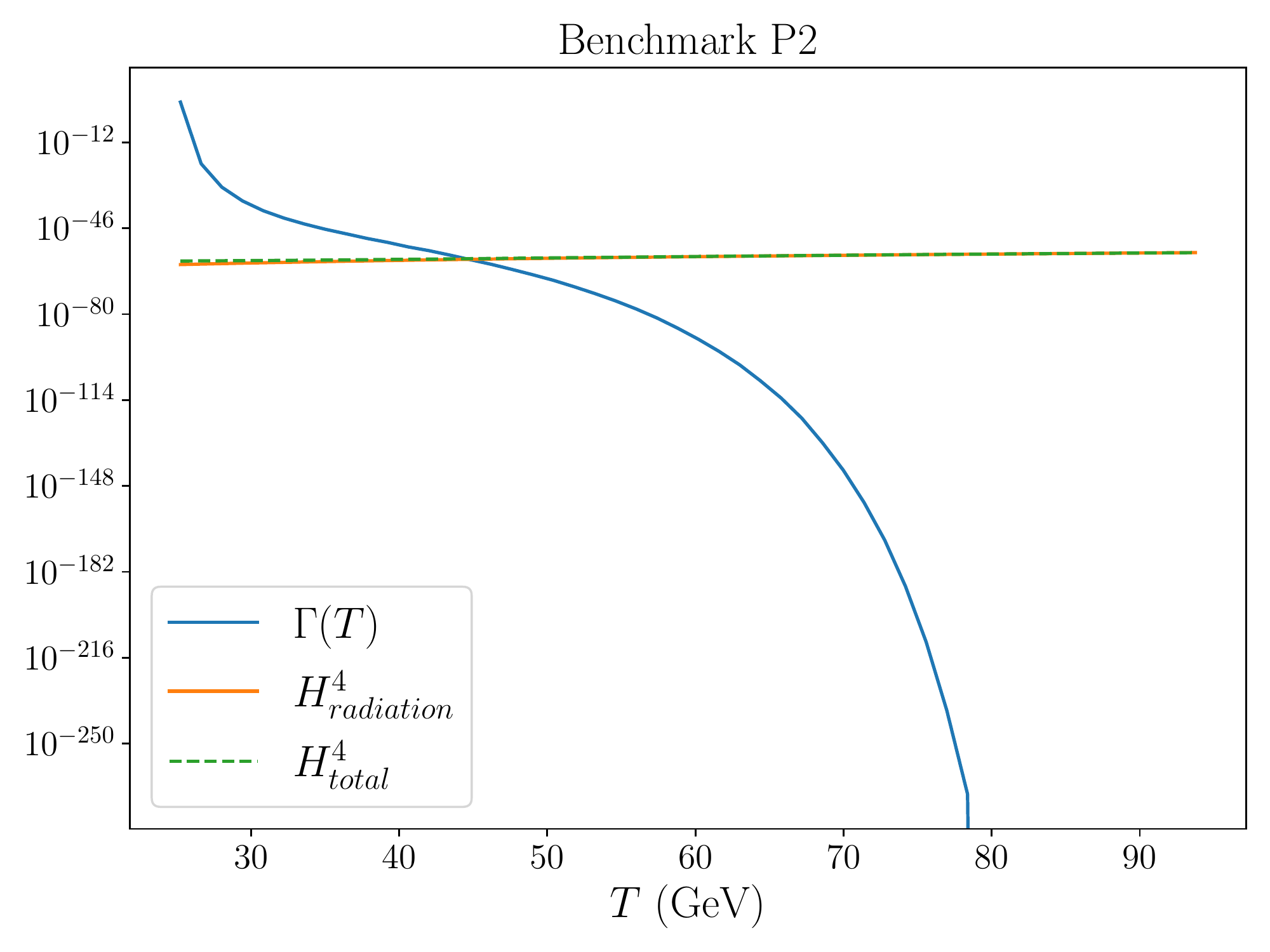}
}\label{Nucleation_assesment}
\caption{\it Rates as functions of temperature for the two benchmarks
introduced in the text. Blue curves show the finite-temperature 
nucleation rates, orange lines the Hubble rates including only the
radiation component and green dashed lines show the total 
Hubble rates with the vacuum contribution included.  }
\end{figure}

To illustrate the importance of a careful nucleation assessment we show in
Fig.~\ref{Nucleation_assesment} the nucleation rate and the Hubble parameter 
evaluated both with only radiation and with the full energy density
including also the vacuum energy-difference term as functions of temperature 
for two benchmark points in the parameter space:  
\begin{align}
    \text{P1}&: \quad m_s=168 \ \text{GeV}, \quad \theta=0.23, \quad u=-148 \ \text{GeV},\quad \mu_{hs}=137 \ \text{GeV}, \quad \mu_3=-577 \ \text{GeV} \, , \nonumber \\
    \text{P2}&: \quad m_s=133 \ \text{GeV}, \quad \theta=-0.02, \quad u=129 \ \text{GeV},\quad \mu_{hs}=-137 \ \text{GeV}, \quad \mu_3=566 \ \text{GeV} \, . \nonumber 
\end{align}
For P1 the vacuum contribution is dominant and prevents nucleation from occurring.
In contrast, for P2 the vacuum contribution is subdominant and we have checked that
bubbles not only nucleate but also satisfy the condition for successful percolation.

\section{Case of Spontaneous $\mathbb{Z}_2$ Symmetry Breaking}\label{sec:spontaneous_Z_2_breaking}

As previously mentioned, the model we are studying can be 
described by five free parameters: $m_s$, $\theta$, $u$, 
$\mu_{hs}$ and $\mu_3$, which are the scalar singlet mass, 
the mixing angle with the SM Higgs boson, the singlet 
vacuum expectation value (vev) and the two $\mathbb{Z}_2$-breaking 
trilinear couplings, respectively. Before presenting the 
full results of our scans, we consider the pattern of the 
phase transitions in simplified scenarios.
We start with the simplest scenario in which the potential is 
symmetric under a parity transformation of the fields. 
This case corresponds to switching off the trilinear couplings, 
i.e., setting $\mu_{hs} = \mu_3 = 0$. The formulae for the quartic
couplings now take the form
\begin{equation}
\begin{split}
\lambda_h &= \frac{1}{2 v^2}(m_h^2\cos^2\theta+m_s^2\sin^2\theta) \, ,  \\
\lambda_s &= \frac{1}{2u^2}(m_h^2\sin^2\theta + m_s^2\cos^2\theta ) \, ,         \\
\lambda_{hs} &= \frac{1}{vu}((m_s^2-m_h^2)\sin\theta\cos\theta ) \, .
\end{split}
\end{equation}
It is important to remark that the parametrization adopted above 
renders it impossible to recover the scenario in which the 
singlet acquires no vev at zero temperature,  
usually referred to as xSM, as the limit $u\rightarrow 0$ is 
ill-defined mathematically. 
In the xSM model, studies have shown~\cite{Lewicki:2021pgr} 
that strong FOPTs are positively correlated with positive values 
of $\lambda_{hs}$. One might be tempted to infer that a 
similar trait also appears in the model under study, 
with the caveat that $\lambda_s$ and $\lambda_{hs}$ 
are no longer independent of each other as in the xSM case.  

The model with a $\mathbb{Z}_2$-symmetric potential was already studied
in~\cite{Carena:2019une} using the leading-order high-temperature
expansion augmented with a trilinear term and neglecting the rest of the Coleman-Weinberg effective potential. 
The authors derived analytical expressions for
relevant quantities at the critical temperature and showed
that the single parameter controlling these quantities is given by
\begin{equation}
    \tilde{\lambda}_h = \lambda_h - \frac{\lambda_{hs}^2}{4\lambda_s} \, .
\end{equation}

\begin{figure}[!h]
\centering
\includegraphics[scale=1]{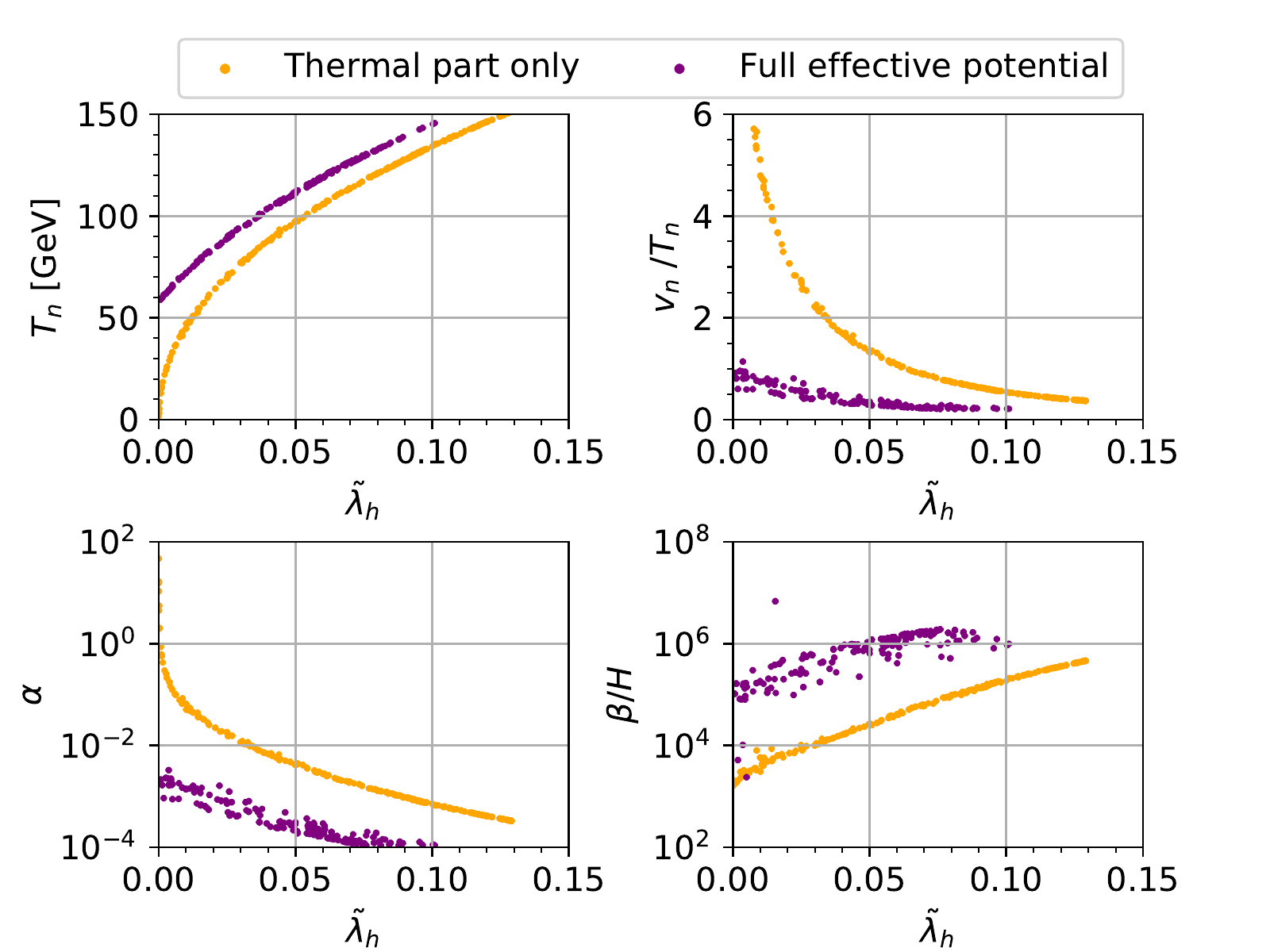}
\label{fig:Z2_comparison}
\caption{\it The phase transition parameters $T_n$, $\frac{v_n}{T_n}$,
$\alpha$ and $\frac{\beta}{H}$ as functions of the effective 
coupling $\tilde{\lambda}_h$ for the thermally-corrected and 
full forms of the effective potential.}
\end{figure}

Here we calculate the parameters of the transition in this model
using the full effective potential. Fig.~\ref{fig:Z2_comparison} 
compares our results with the approximate results found in~\cite{Carena:2019une}.
In general, a correlation with the effective coupling 
$\tilde{\lambda}_h$ is still visible, but 
transitions using the full potential are significantly weaker. 
As $\tilde{\lambda}_h$ approaches $\lambda_h$, 
transitions become second-order and the trilinear coefficient $E$
acquires its SM value. However, in the limit of stronger transitions
tunnelling becomes singlet-driven, so the impact of the singlet 
becomes important, leading to vanishing $E$. Hence, for the 
strongest transitions possible in this model, the existence of the
barrier is ensured at the tree level. In summary, we conclude that
the $\mathbb{Z}_2$-symmetric version of the theory predicts only very weak transitions and cannot provide any significant 
observational predictions for GWs or baryogenesis. Therefore,
we do not discuss this scenario further in the subsequent Sections of this 
paper.~\footnote{All our calculations were performed using the
on-shell renormalization scheme, but we have verified that the results 
obtained using the MS-bar renormalization scheme are almost identical.}

\section{Properties of the Bubble Wall} \label{sec:wall_properties}

Estimating the properties of the bubble wall of a FOPT entails an involved
computation of out-of-equilibrium perturbations in the plasma. In addition, 
the form of the  equations that should be used to obtain these estimates 
as well as the assumptions underlying them are not agreed upon within the 
particle physics community. The most common presumptions are (i) that the 
bubble wall and the plasma particles interact in local thermal 
equilibrium \cite{Konstandin:2010dm, BarrosoMancha:2020fay, Balaji:2020yrx, Ai:2021kak, Wang:2022txy} or alternatively, 
(ii) that only the heaviest particles are 
taken out of thermal equilibrium by their interaction with the advancing wall
but the rest of the particles in the plasma are treated as a background fluid in
thermal equilibrium \cite{Dine:1992wr,Liu:1992tn, Bodeker:2009qy, Bodeker:2017cim},
henceforth called the \textit{semiclassical fluid approximation}. In this paper 
we advocate the latter possibility and use the modified fluid equations of 
Refs.~\cite{Laurent:2020gpg, Cline:2021iff}. For alternative developments 
regarding assumption (ii) see \cite{Laurent:2022jrs, Dorsch:2021nje, DeCurtis:2022hlx}.

The dynamics of the wall in the fluid approximation were first estimated in the
pioneering papers \cite{Moore:1995si,Moore:1995ua} for the SM electroweak theory
with a light Higgs boson mass. Subsequently, the discovery of the Higgs boson prompted the adoption of this approximation in phenomenological models 
with a strong FOPT and not a crossover. The preferred model for study has been 
the scalar gauge singlet extension, either with a $\mathbb{Z}_2$-symmetric potential 
and no vev at zero temperature,
see~\cite{Cline:2021iff,Laurent:2022jrs,Friedlander:2020tnq} or without the $\mathbb{Z}_2$
symmetry and non-vanishing vev at zero temperature~\cite{Kozaczuk:2015owa}.
Additionally, the SM effective field theory (SMEFT) with dimension-6 operator 
and a low cutoff was investigated in \cite{Konstandin:2014zta, Huber:2013kj}.  

We studied in \cite{Lewicki:2021pgr} the generic implications that the  
\textit{new formalism} (as dubbed in \cite{DeCurtis:2022hlx}) of
\cite{Laurent:2020gpg, Cline:2021iff} has for the bubble wall properties 
in the $\mathbb{Z}_2$-symmetric singlet model with vanishing vev at zero temperature 
and in the SMEFT. One of the central results in \cite{Lewicki:2021pgr} 
was that one can obtain a good numerical approximation for the wall speed 
using thermal equilibrium as a starting assumption. The following 
analytic formula was derived:
\be \label{eqn:approx_velocity}
v_{\text{approx}}=
\begin{cases}
\sqrt{\frac{\Delta V}{\alpha \rho_r}} \quad \quad {\rm for} \quad \sqrt{\frac{\Delta V}{\alpha \rho_r}}<v_J(\alpha),
\\
1 \quad \quad \quad \quad {\rm for} \quad  \sqrt{\frac{\Delta V}{\alpha \rho_r}} \geq v_J(\alpha) \, ,
\end{cases}
\ee
in which $\Delta V $ is the potential difference between the false and true vacua evaluated at the nucleation temperature, $\alpha \rho_r$ denotes the latent heat released and 
\be \label{eq:vJ}
v_J=\frac{1}{\sqrt{3}}\frac{1+\sqrt{3 \alpha^2+2 \alpha}}{1+\alpha} \, ,
\ee
is the Chapman-Jouguet velocity~\cite{Steinhardt:1981ct,Kamionkowski:1993fg,Espinosa:2010hh}.
The analytic approach was compared in~\cite{Lewicki:2021pgr} to the full
computation for the two aforementioned BSM scenarios. The predictions
of the two calculations were shown to agree well. This result is linked 
to the fact that the deviations from equilibrium are small and one can 
neglect temperature variations in the derivation of the analytic formula. 
This has been verified in \cite{Laurent:2022jrs}, in which it was found that
deviations from equilibrium are subdominant and that for sufficiently
weak transitions a hydrodynamic treatment of the plasma shows that
the bubble wall can reach a steady state.

In this paper we carry this analysis over to study the scalar singlet 
without $\mathbb{Z}_2$ symmetry and non-vanishing zero temperature vev.
To the best of our knowledge, the computation of the bubble wall properties in this 
scenario has only been studied in Ref. \cite{Kozaczuk:2015owa} using, however, 
the fluid equations laid out in \cite{Moore:1995si,Moore:1995ua} that yield 
singularities for values of the wall velocity close to the speed of sound in the plasma. 
The computations in our work avoid this problem, and so
represent a significantly improved assessment of the status 
of the model as a viable framework for electroweak baryogenesis.

With regards to the algorithm used in \cite{Lewicki:2021pgr}, 
the scenario studied here requires only a minor modification. 
Clearly, the friction on the bubble wall in both the $\mathbb{Z}_2$-symmetric and 
non-symmetric cases is the same. Thus we follow the procedure of considering 
only the dominant contributions to the friction due to the 
electroweak gauge bosons and the top quark. The other particles, 
including the Higgs boson and singlet field, are treated as a 
background perturbations with zero chemical potential. 

The necessary amendment for this scenario is due to the different form of 
scalar potential. In particular, the vacuum structure and the larger number of 
free parameters produce different features in the phase transition. 
For most values of the free parameters, the pattern of the transition starts from a 
high-temperature phase with zero Higgs vev and non-zero singlet vev
(either positive or negative), and then tunnels towards the global minimum 
of the theory. As the field profiles should interpolate between the false 
and true vacua, we replace Eqns.~(4.14) and (4.15) in
Ref.~\cite{Lewicki:2021pgr} by the modified Ansatz 
\be
h(z) = \frac{h_0}{2 }\left[ \tanh{\left( \frac{z}{L_h} \right)} + 1\right],
\ee
\be
s(z) = \frac{s_l-s_h}{2 } \tanh{\left( \frac{z}{L_s} - \delta_s \right)}  + \frac{s_h+s_l}{2} \, ,
\ee
where $h_0$ is the Higgs vev at the true minimum and $s_h$ and $s_l$ 
represent the high-temperature (metastable) and low-temperature (stable)
vevs of the singlet. The rest of the algorithm used in Ref.~\cite{Lewicki:2021pgr}
remains unchanged. 

\begin{figure}[h]
\includegraphics[scale=.67]{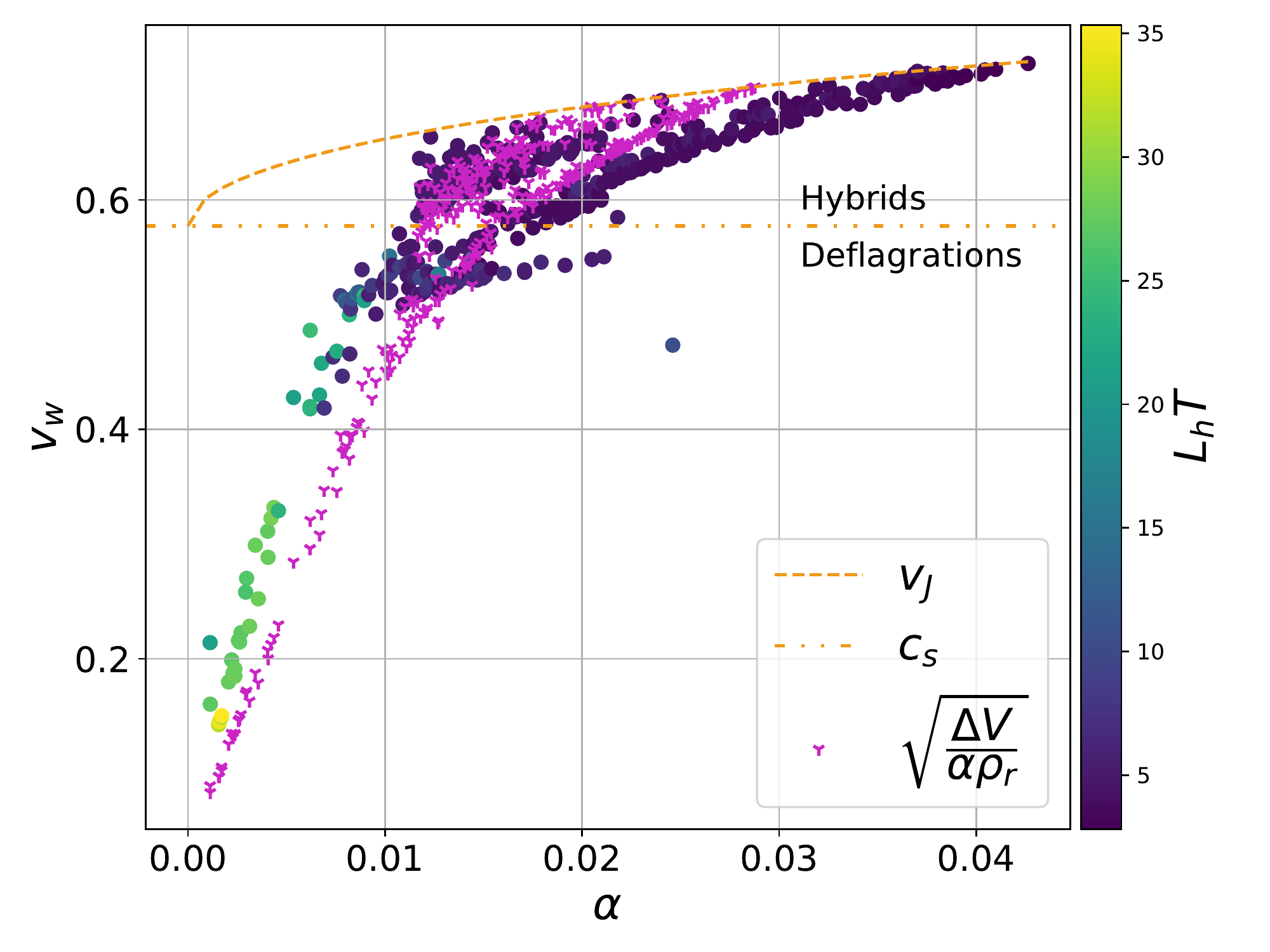} 
\centering
\caption{\it Predictions for the bubble wall velocity as a function
of the strength $\alpha$ of the phase transition. The coloured side-bar gives 
the value of the Higgs profile thickness normalized to the nucleation temperature. 
The magenta crosses represent the values of the analytic approximation. 
The  orange dashed and dash-dot-dotted lines represent the Chapman-Jouguet velocity 
and the speed of sound in the plasma, respectively.}
\label{singlet_bubble}
\end{figure}

Results from the computation of the bubble wall velocity for a random scan 
of the parameter space are shown in Fig.~\ref{singlet_bubble}, where
the wall velocity is plotted against the strength of the transition
and the coloured side-bar indicates the bubble wall width.
The dash-dot-dotted orange line is the speed of sound in the plasma, 
$c_s =1/\sqrt{3}$, below which the explosive growth of the bubble wall is a deflagration. 
The dashed orange line shows the Chapman-Jouguet velocity at a given value of $\alpha$,
while the magenta crosses are the value of the wall velocity obtained using the 
analytical approximation. In the region between the Chapman-Jouguet velocity and the 
speed of sound the explosive advancement of the bubble is a hybrid, 
i.e., it is composed of a shock discontinuity in front of the wall 
and a rarefaction wave behind it. 

As shown in the figure, we cannot find steady-state solutions
for points above the Chapman-Jouguet velocity. 
In those cases we expect the wall to reach 
highly relativistic velocities $v_w \approx 1$ and the 
expansion to proceed as a detonation. For some points of the 
random scan the analytic formula
(corresponding to the magenta crosses) predicts $v_w=1$ 
whereas in fact a 
steady-state solution can be obtained. 

\begin{figure}[!h]
\centering
\subfigure{
\includegraphics[width=7.2cm]{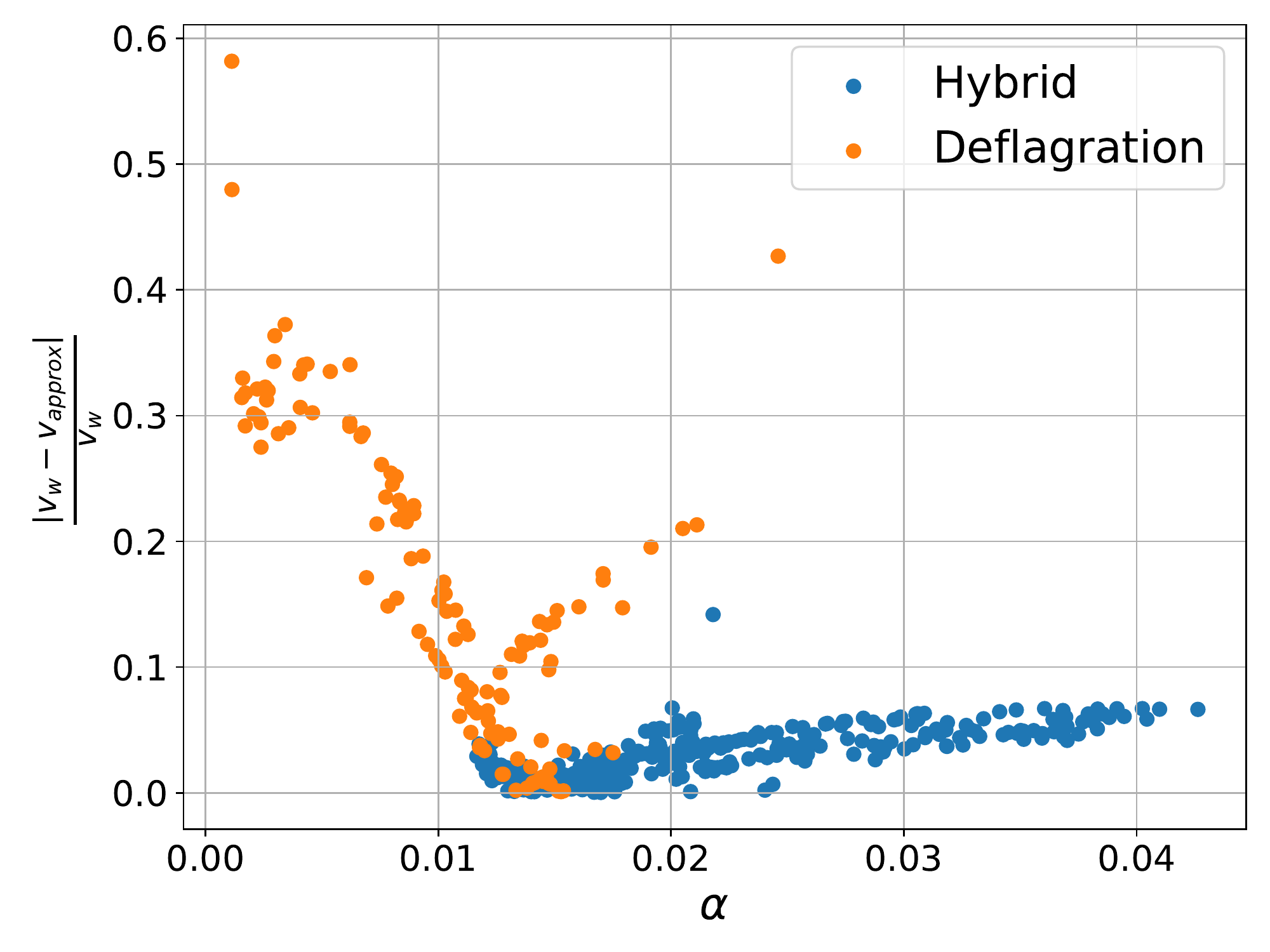}
}
\subfigure{
\includegraphics[width=7.2cm]{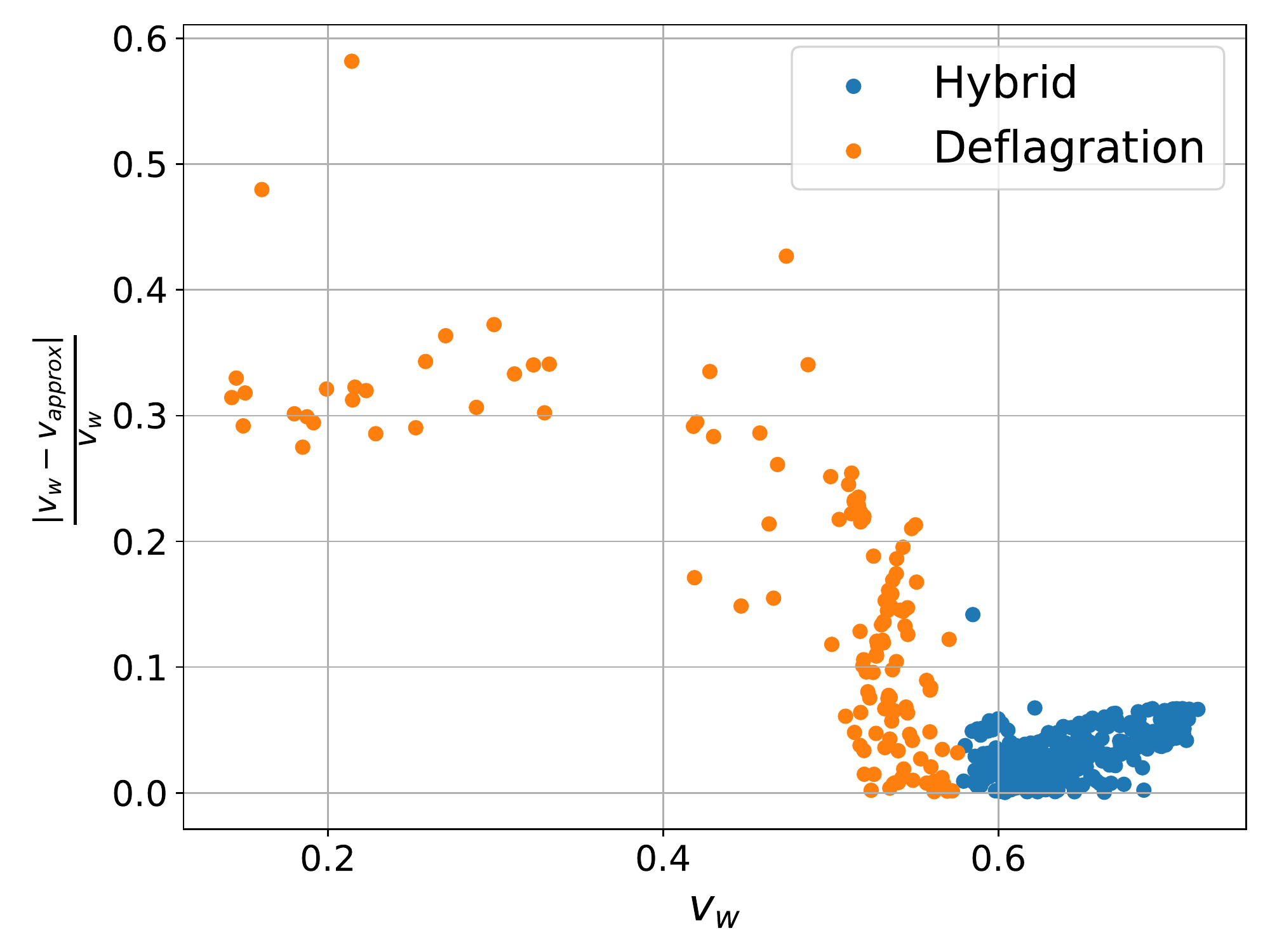}
}
\caption{\it Relative errors in the wall velocity between the analytic approximation and the full calculation for deflagration points (orange) and hybrids (blue).}
\label{relative_error}
\end{figure}

In order to quantify the applicability of the analytic approximation,
we computed the sample mean percentage error between the analytic formula 
and the full computation:
\begin{equation}
 \bar{v} \equiv \frac{1}{N} \sum_i^{N} \frac{|v_w - v_{\text{approx}}|}{v_w}\times 100 \approx 7.1 \% \, , 
\end{equation}
which shows that the approximation provides a remarkably good estimate. 
The relative error for each parameter space point is shown in Fig.~\ref{relative_error}. 
The expansion profiles for deflagrations (hybrids) are shown as orange (blue). 
We find that the relative error is less than about ten percent for hybrid points,
whereas deflagrations exhibit stronger deviations.

\begin{figure}[!h]
\centering
\subfigure{
\includegraphics[scale=.47]{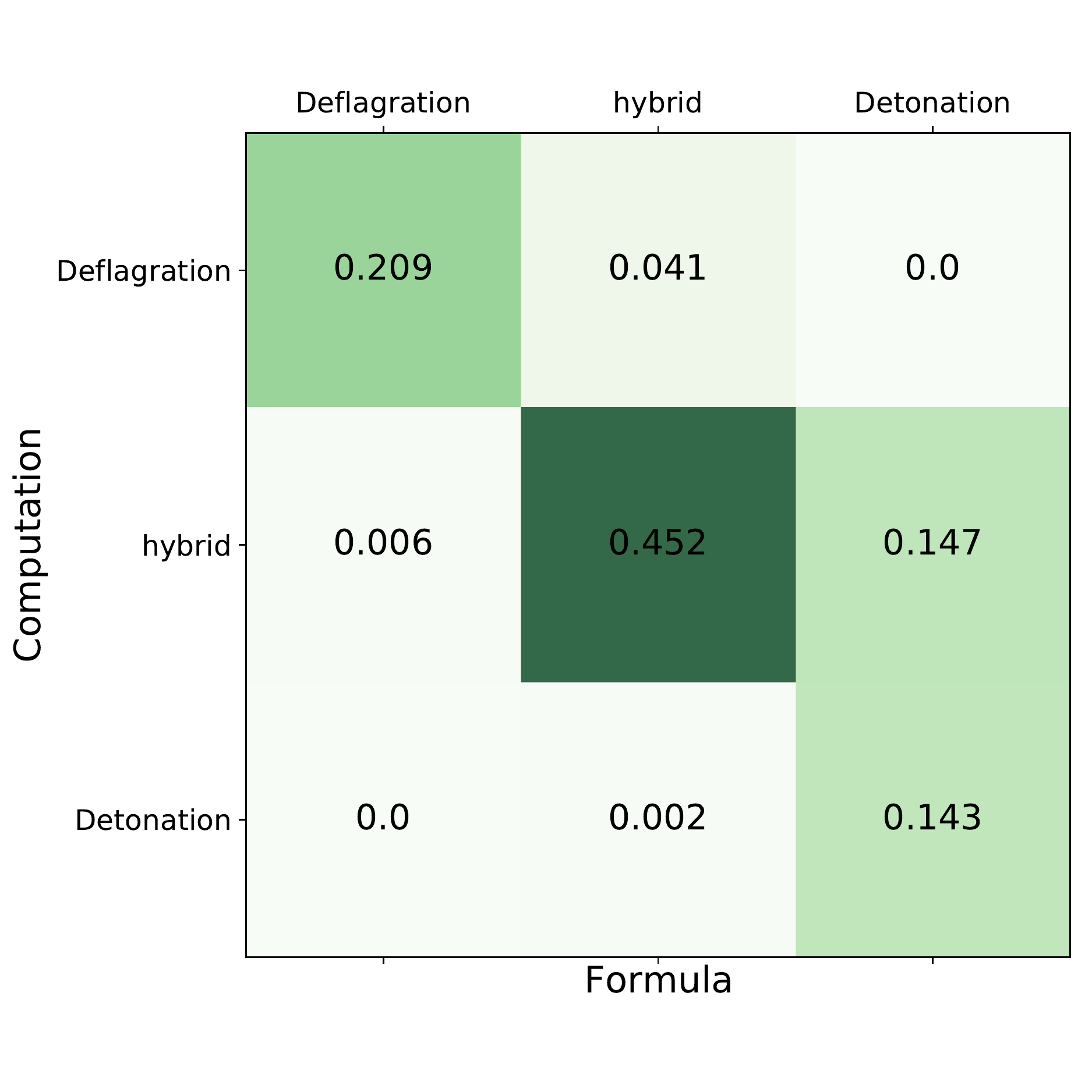}
}
\caption{\it The confusion matrix for multi-class classification of the hydrodynamic
expansion based on the computation of the wall velocity~\cite{scikit-learn}. 
The rows (columns) denote the
fraction of points that fall within each class as obtained from a full numerical
computation (analytic formula). }
\label{confusion_matrix}
\end{figure}

In order to assess further the utility of the analytic formula for
classifying the type of hydrodynamic expansion, we have made a 
confusion matrix analysis whose results are shown in Fig.~\ref{confusion_matrix}. 
The rows of the matrix correspond to the outcome of the full computation,
whereas the columns give the prediction of the analytic formula.
The last row corresponds to points for which a steady-state solution was not found. 
The numerical values of each cell are normalized to the total number of points in the scan. 

In the language of the confusion matrix for multi-class classification,
the diagonal elements
represent the fraction of times the analytic prediction classified the solution correctly,
and the off-diagonal matrix elements the fraction of times it misclassified them. We can see that the formula classified incorrectly about $15\%$ of the points as detonations when in fact they were hybrids, see the 23 matrix element. 
We made use of the {\tt scikit-learn} python package~\cite{scikit-learn} 
for the evaluation of the confusion matrix elements and obtained a 
total weighted F-score of $F_1 = 0.8$, which is obtained from the harmonic mean of the precision. We recall that the best possible F-value is $1$ and poorest F-value is $0$. 

\newpage

\section{Gravitational Waves}\label{sec:GWs}
The computation of the gravitational wave spectrum relies on estimates for the thermodynamic parameters of the FOPT. 
One of these key parameters, the percolation temperature, was already discussed in the previous Section, and we introduce the three remaining parameters in this Section.

The strength of the phase transition is proportional to the trace anomaly, i.e.,
\be
\alpha \equiv  \frac{1}{4 \rho_r} \Delta\Tr{ T^{\mu}_{\nu}}   = \frac{1}{\rho_r}\left( \Delta V_{eff}(\phi,T) - \frac{T}{4} \Delta \frac{\partial V_{eff}(\phi,T)}{\partial T} \right) \, , \label{alpha_def}
\ee
where $\Delta$ denotes the differences between 
quantities in the false and true minima. 
Another relevant parameter, denoted by $\beta$, defined by
\be \label{eq:betaH}
\frac{\beta}{H} \equiv T_* \frac{d}{dT} \left(  \frac{S_3}{T} \right)\Big{|}_{T=T_*}\, ,
\ee 
specifies the inverse time duration of the phase transition.
We evaluate the GW spectrum at the temperature reached after the transition when the vacuum energy is converted into radiation:
\be
T_* = T_p(1+\alpha(T_p))^{1/4} \, .
\ee
The final relevant parameter is the wall velocity $v_w$, which we discussed in detail in Sec~\ref{sec:wall_properties}. Here we will use the results of our detailed numerical analysis although, as we have shown, the analytic estimate from Eq.~(\ref{eqn:approx_velocity}) would yield very similar results.


Since the potential is polynomial, we do not expect significant
supercooling~\cite{Ellis:2018mja}, which implies that the bubbles will not become very
energetic~\cite{Ellis:2019oqb,Lewicki:2019gmv,Lewicki:2020jiv,Lewicki:2020azd,Lewicki:2022pdb}.
Hence we can focus exclusively on GWs sourced by plasma
motion~\cite{Caprini:2019egz,Caprini:2015zlo}. 
Despite significant progress in the modelling of
turbulence~\cite{RoperPol:2019wvy,Kahniashvili:2020jgm,Pol:2021uol,Auclair:2022jod}, the calculation of GWs from its emergence after a phase transition remains uncertain,
and we will omit this source.
This leaves us with sound waves as the main source, for which we use the 
results of lattice simulations to compute the GW
signal~\cite{Hindmarsh:2013xza,Hindmarsh:2015qta,Hindmarsh:2017gnf,Hindmarsh:2020hop}:
\bea
&\Omega_{\rm sw}(f)h^2 = 4.13\times 10^{-7} \, \left(R_* H_*\right)  \left(1- \frac{1}{\sqrt{1+2\tau_{\rm sw}H_*}} \right)  \left(\frac{\kappa_{\rm sw} \,\alpha }{1+\alpha }\right)^2 \left(\frac{100}{g_*}\right)^\frac13 S_{\rm sw}(f) \,, \\
{\rm where} \; & S_{\rm sw}(f)=\left(\frac{f}{f_{\rm sw}}\right)^3 \left[\frac{4}{7}+\frac{3}{7} \left(\frac{f}{f_{\rm sw}}\right)^2\right]^{-\frac72} \,,
\eea
the frequency of the peak is given by
\begin{equation}
 f_{\rm sw} \,=2.6\times 10^{-5} {\rm Hz} \left(R_* H_*\right)^{-1} \left(\frac{T_*}{100 {\rm GeV}}\right)\left(\frac{g_*}{100}\right)^\frac16 \,.  
\end{equation}
and $g_*$ is the number of degrees of freedom at $T_*$, which we compute using the results of~\cite{Saikawa:2018rcs}.
In order to approximate the duration of the sound wave period normalised to the Hubble rate we use~\cite{Hindmarsh:2017gnf,Ellis:2018mja,Ellis:2019oqb,Ellis:2020awk,Guo:2020grp}
\begin{equation}
\tau_{\rm sw}H_* =\frac{H_* R_*}{U_f}\,, \quad U_f\approx \sqrt{\frac34 \frac{\alpha}{1+\alpha} \kappa_{\rm sw}}\,,
\end{equation}
and to approximate the average bubble radius, again normalised to the Hubble rate, we use
\begin{equation}
H_*R_* \approx (8\pi)^\frac13 \, {\rm Max}(v_w,c_s)\left(\frac{\beta}{H}\right)^{-1} \, ,
\end{equation}
where we use the duration of the transition given in
Eq.~\eqref{eq:betaH}.
Finally, we use the fluid velocity and temperature profiles 
discussed in Section~3 of Ref. \cite{Lewicki:2021pgr} to calculate the sound wave efficiency factor as the energy converted into 
bulk fluid motion, which is given by~\cite{Espinosa:2010hh}
\be
\kappa_{sw} = \frac{3}{\alpha\, \rho_{\mathrm{R}} \,v_w^3} \int w\, \xi^2 \frac{v^2}{1-v^2} d\xi= \frac{4}{\alpha \,v_w^3} \int \left( \frac{T(\xi)}{T_p} \right)^4\, \xi^2 \frac{v^2}{1-v^2} d\xi \, .
\label{eq:kappa_eff}
\ee 
The signal-to-noise ration observed by a given experiment is given by 
\begin{equation}
    \text{SNR} \equiv \sqrt{T} \left[   \int_{f_{min}}^{f_{max}} \left( \frac{\Omega_{\text{sw}}(f)}{\Omega_{\text{noise}}(f)}  \right)^2  df\right]^{1/2} \, ,
\end{equation}
which we calculate assuming the duration of each mission to be $T=4$~years. We visualise the sensitivities of experiments using the standard power-law integrated sensitivities~\cite{Thrane:2013oya}.

\begin{figure}[!h]
\centering
\includegraphics[width=14cm]{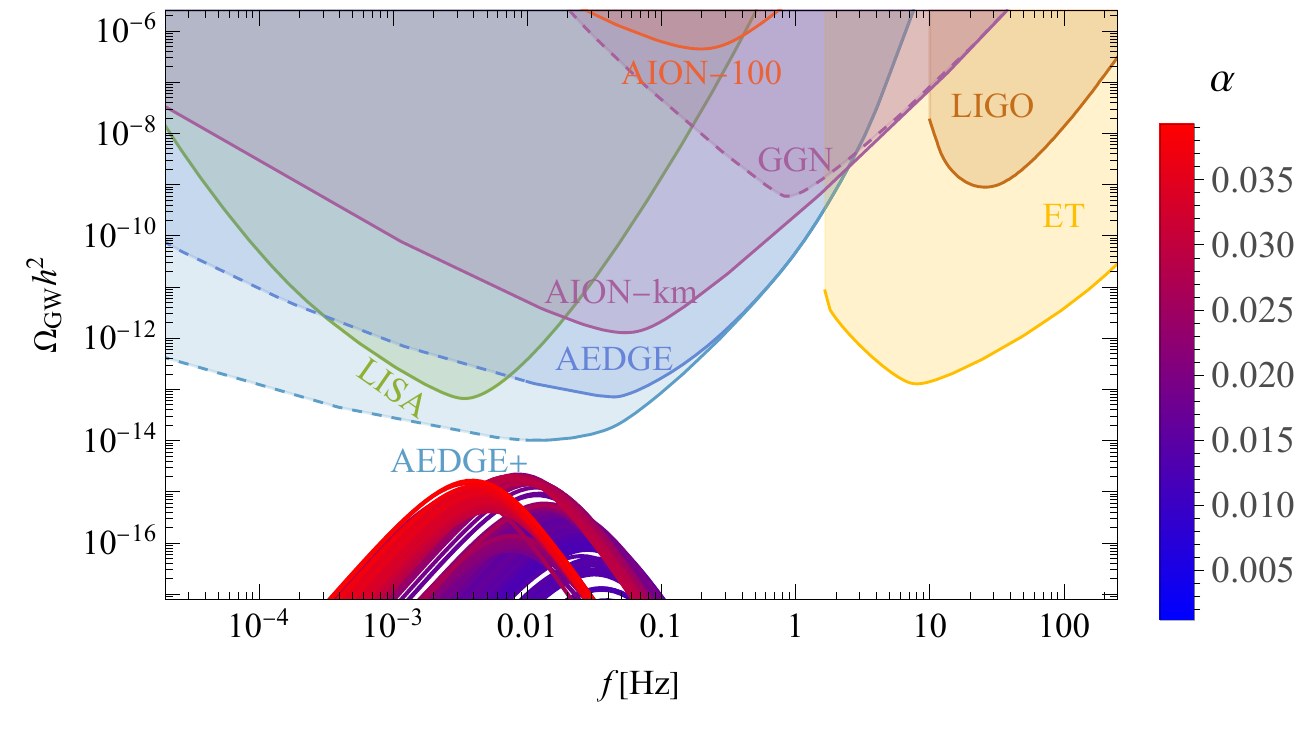}
\vspace{-5mm}
\caption{\it Gravitational wave spectra from points with bubble walls
that are not very relativistic, for which we find hydrodynamical solutions.}
\label{fig:GW_SlowWall}
\end{figure}

\begin{figure}[!h]
\centering
\includegraphics[width=14cm]{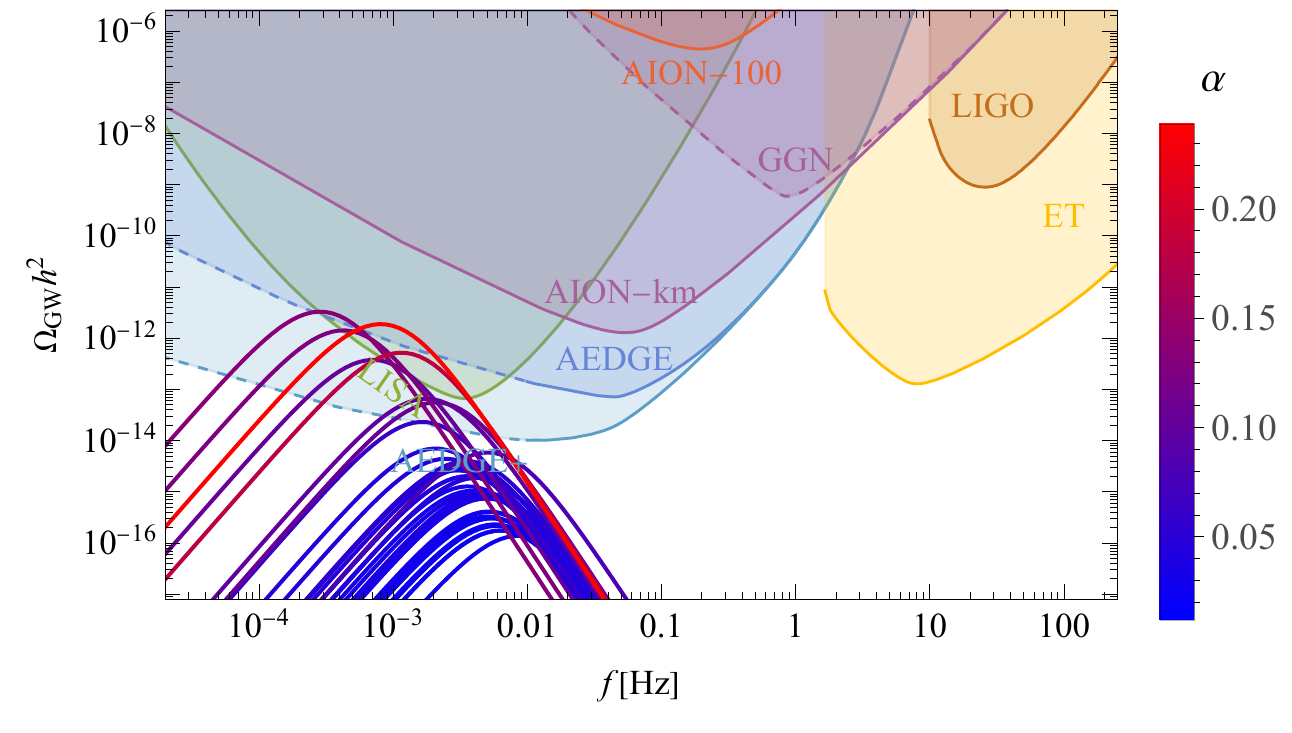}
\caption{\it Gravitational wave spectra from points in the parameter space for which the transitions are too strong and we do not find hydrodynamical solutions. Here we simply assume $v_w\approx 1$. We note that these points are not suitable for baryogenesis.} 
\label{fig:GW_FastWalls}\vspace{5mm}
\end{figure}

We show the spectra from our scans in Figs.~\ref{fig:GW_SlowWall} and \ref{fig:GW_FastWalls}, with the first one showing only points for which we were able to compute the wall profile and velocity, while in the second one the friction was not enough for the walls to reach a steady state and we assumed $v_w\approx1$ instead. Both figures also show the design sensitivity of the currently running LIGO~\cite{LIGOScientific:2014pky,Thrane:2013oya,LIGOScientific:2016fpe,LIGOScientific:2019vic} and future interferometers LISA~\cite{Bartolo:2016ami,Caprini:2019pxz} and
ET~\cite{Punturo:2010zz,Hild:2010id} as well as proposed devices based on atom interferometry~\cite{Badurina:2021rgt}, AION-1km~\cite{Badurina:2019hst} and AEDGE~\cite{AEDGE:2019nxb}.
We do not include the impact of the GW foreground produced by the population of BH currently probed by LIGO and Virgo~~\cite{Lewicki:2021kmu}. This effect would likely further limit our detection prospects, however, it would not change our conclusions.
We see only points where the wall velocity is close to unity and do not find hydrodynamical solutions that produce signals strong enough to be observed in upcoming experiments. This trend follows similar observations made in the SMEFT and the $\mathbb{Z}_2$-symmetric versions of the model~\cite{Cline:2021iff,Lewicki:2021pgr}.

Significant progress has been made recently on hybrid calculations of GW generation
by sound waves, which shows non-trivial dependence of the shape of the spectrum on the
wall velocity~\cite{Hindmarsh:2016lnk,Hindmarsh:2019phv,Jinno:2020eqg,Gowling:2021gcy}.
However, these modifications do not have a large impact on the peak GW density. They
would only modify the spectra in Fig.~\ref{fig:GW_SlowWall} when the velocities are
non-relativistic. As the spectra in such cases are too weak to be observed in future
experiments, the overall impact of these updates on upcoming searches are not expected to be very significant in the model 
studied here.





\section{Baryogenesis}
\label{sec:Baryogenesis}


We discuss in this Section the use of the CP-violating dimension-$5$ operator
\begin{equation}
    \mathcal{L} \supseteq y_t \bar{Q} \tilde{\Phi} t_R \left(i\frac{s}{\Lambda_{CP}} \right)  + \text{h.c.}
    \label{eq:dim_5_Operator}
\end{equation}
to achieve baryogenesis in the general singlet extension of the SM discussed in Section~\ref{sec:Veff}.~\footnote{A 
related dimension-$6$ six operator, i.e., with $s/\Lambda_{CP} \to s^2/\Lambda_{CP}^2$, was considered in
\cite{Vaskonen:2016yiu,Cline:2012hg} as a possible term in a $\mathbb{Z}_2$-symmetric theory.} 
The baryon asymmetry that can be generated by this operator
has been investigated previously, mostly in the singlet extension 
with a $\mathbb{Z}_2$-symmetric potential and vanishing vev \cite{Cline:2021iff, Lewicki:2021pgr}.
Here we first review
the phenomenological constraints on this model before exploring the possible magnitude of the baryon asymmetry that
it can generate in the presence of a zero-temperature vev $u$, under various conditions. 

\subsection{Higgs signal strengths and CP properties}
\label{Sec:pheno}

In the presence of the  
operator \eqref{eq:dim_5_Operator}, the top quark mass deviates from the SM expression $m_t=y_t^{\text{SM}} v/\sqrt{2}$ when the singlet develops a zero-temperature vev $u$. Hence the top-quark Yukawa coupling $y_t$ must be rescaled accordingly:
\begin{equation}
    y_t = \frac{ y_t^{\rm SM}}{ \sqrt{1 +  \frac{u^2}{\Lambda^2_{CP}}}} = \frac{m_t\, \sqrt{2}}{ v\, \sqrt{1 +  \frac{u^2}{\Lambda^2_{CP}}}}
\end{equation}
in order to reproduce correctly the top-quark mass value obtained from Tevatron and LHC measurements (see, e.g.,~\cite{ATLAS:2014wva}), for which we use the Particle Data Group value $m_t = 172.9$ GeV~\cite{ParticleDataGroup:2020ssz}.
%
%

\vspace{1mm}

\begin{figure}[hbt!]
\centering
\subfigure{
\includegraphics[width=0.485\textwidth]{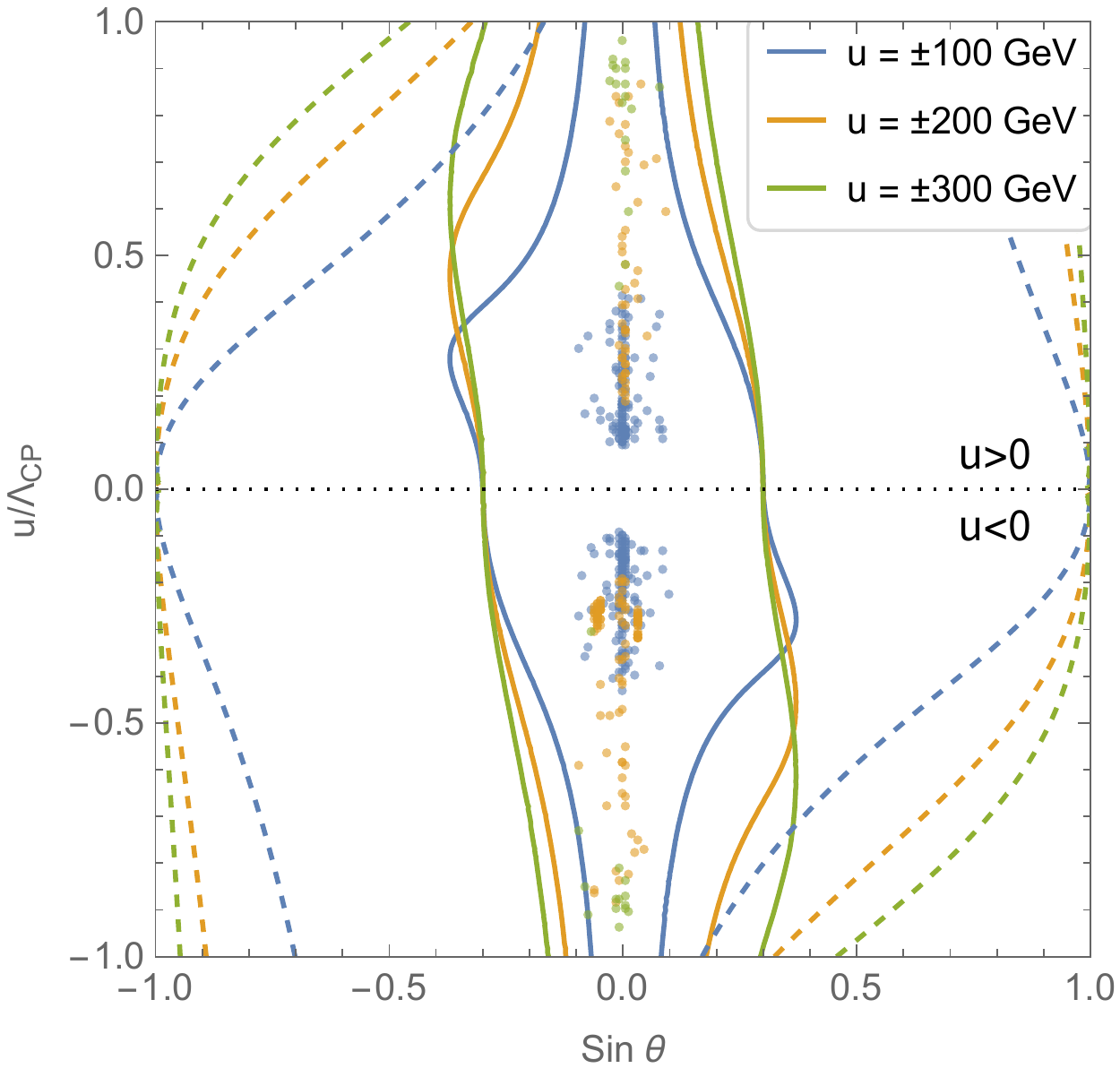}
\hspace{1mm}
\includegraphics[width=0.485\textwidth]{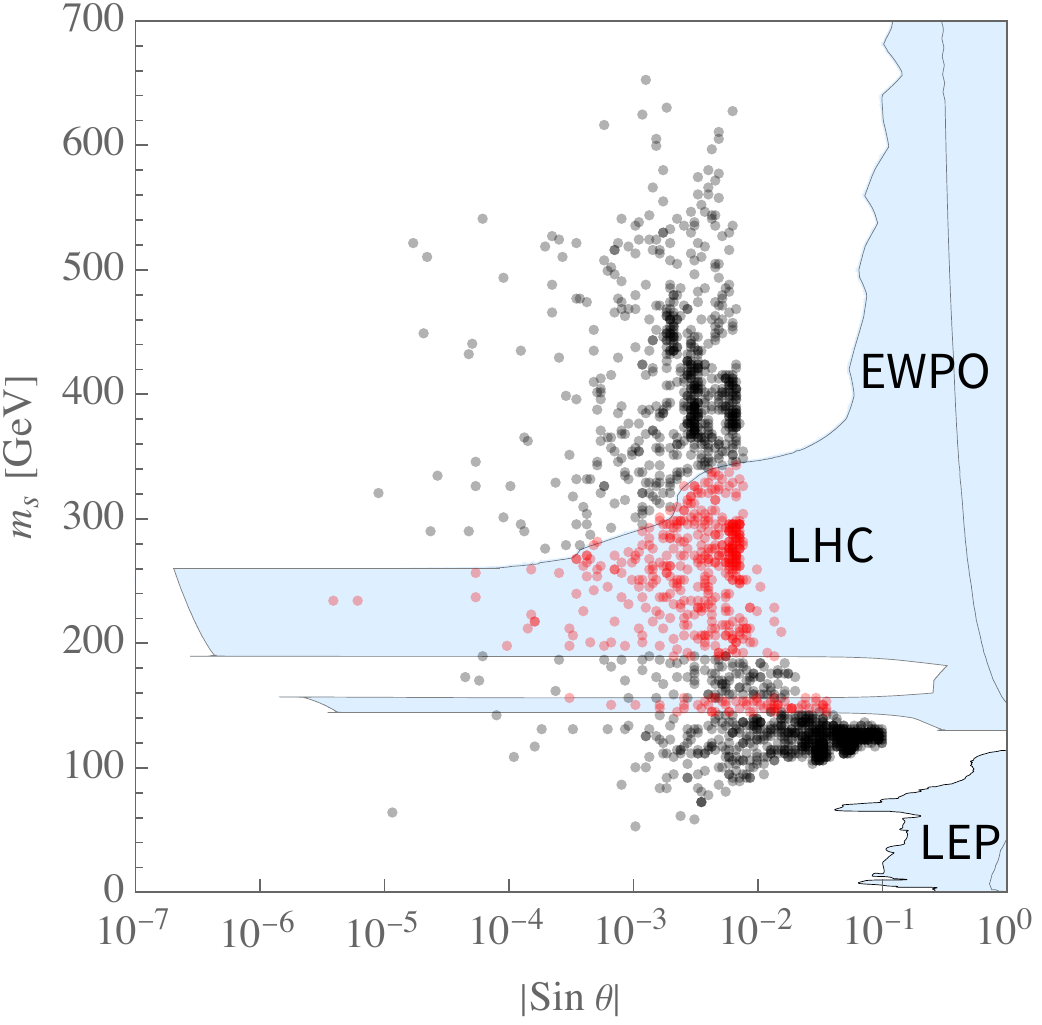}
}
\caption{\it Left: 2-$\sigma$ allowed regions in the $(\sin \theta, u/\Lambda_{CP})$ plane for different values of the singlet vev $u$, from a $\chi^2$ fit to the Higgs signal strength measurements from ATLAS in its 10$^{\rm th}$-year Higgs Legacy paper~\cite{ATLAS:2022vkf} (solid lines), and from a measurement of the CP structure of the $h \bar t t$ coupling by CMS~\cite{CMS:2020cga} (dashed lines). Right: The $(\sin \theta, m_s)$ plane showing regions excluded at the 2-$\sigma$ level by LEP searches for light singlet-like scalars, LHC searches, and measurements of EW precision observables.
The points selected for our baryogenesis analysis in Section~\ref{sec:BAU} are shown in black on the right panel (the red points are excluded by LHC searches), and in colours corresponding to the indicated values of $u\pm 5\%$ in the left panel.} 

\label{ChiSquare_Higgs}
\end{figure}

In parallel, singlet-doublet mixing through the angle $\theta$~\footnote{Hereafter we use the notations $h$ and $s$ for the mass eigenstates that are
predominantly Higgs and singlet, respectively.} reduces universally the couplings of the Higgs boson to SM particles by a factor 
cos $\theta \equiv c_{\theta}$. In the specific case of the top quark, this reduction combines with the modification 
of the Higgs-top coupling induced by the effective operator \eqref{eq:dim_5_Operator} 
to yield a Higgs-top interaction 
$(m_t/v)\, h \,\bar{t}\, (\kappa_t + i\,\tilde{\kappa}_t\, \gamma_5)\, t $, where
\begin{equation}
\kappa_t =  \frac{ c_{\theta}}{ \sqrt{1 +  \frac{u^2}{\Lambda^2_{CP}}}} \quad , \quad 
\tilde{\kappa}_t = \frac{ c_{\theta} \frac{u}{\Lambda_{CP}}  -  s_{\theta}  \frac{v}{\Lambda_{CP}} }{ \sqrt{1 +  \frac{u^2}{\Lambda^2_{CP}}}}\, .
\end{equation}
The corresponding coupling modifier w.r.t.~the SM value is $\left|g_{htt} \right| = \sqrt{\kappa_t^2 + \tilde{\kappa}_t^2 }$.
%
%
%
%
These effects impact the predictions for Higgs signal strength measurements at the LHC and their interpretation.\footnote{We 
do not discuss here constraints from Higgs self-coupling measurements, which are currently relatively weak.}
We have performed a $\chi^2$ fit to the latest Higgs signal strength
measurements published by ATLAS in its 10$^{\rm th}$-year Higgs Legacy analysis~\cite{ATLAS:2022vkf},\footnote{The corresponding Higgs signal strength
analysis by CMS~\cite{CMS:2022dwd} cannot be reinterpreted in our framework, so we do not include it in our $\chi^2$ fit.} in terms of the parameters $s_{\theta}$ and $u/\Lambda_{CP}$. 
The $2\,\sigma$ ($\Delta \chi^2 = 6.18$) allowed regions for different values of
the singlet vev $u$ are enclosed by the solid lines in the left panel of Fig.~\ref{ChiSquare_Higgs}, distinguishing the cases of positive and negative $u$.
In addition, a recent CMS analysis has constrained the CP structure of the Higgs-top interaction~\cite{CMS:2020cga}, yielding the $2\,\sigma$ bound $\tilde{\kappa}_t^2/(\kappa_t^2 + \tilde{\kappa}_t^2) < 0.669$, and the corresponding allowed region of the 
($s_{\theta}$, $u/\Lambda_{CP}$) plane lies within the dashed contours in Fig.~\ref{ChiSquare_Higgs} (left). However, we find that
this bound on the Higgs CP properties is weaker that from Higgs signal strength measurements throughout the entire parameter space of the model. The coloured points in Fig.~\ref{ChiSquare_Higgs} (left) are those selected for our baryogenesis analysis in Section~\ref{sec:BAU}, and we see that they are all comfortably consistent with the constraints from LHC Higgs Signal Strengths.

\subsection{Electroweak precision observables}
\label{sec:EWPO}

In general, the extra scalar in our model makes corrections to the gauge boson self-energy diagrams. The leading effects on the electroweak precision observables (EWPOs) are described by the oblique parameters $S, T$ and $U$. Since the new scalar is electrically neutral, only the $W$ and $Z$ boson self-energies receive corrections.
Complete expressions for the shifts in the oblique parameters from their SM values,
$\Delta S, \Delta T$ and $\Delta U$, are given in~\cite{Beniwal:2018hyi}. 
The respective numerical values of $\Delta S, \Delta T$ and $\Delta U$ obtained from a global fit to EWPO measurements~\cite{Haller:2018nnx} are
\begin{equation}  
\Delta S \; = \; 0.04 \pm 0.11, \; \Delta T \; = \; 0.09 \pm 0.14, \; \Delta U = -0.02 \pm 0.11 \, ,
\label{DeltaObliques}
\end{equation}
together with the correlation coefficients $+0.92$ between $\Delta S$ and $\Delta T$, $-0.68$ between $\Delta S$ and $\Delta U$ and $-0.87$ between $\Delta T$ and $\Delta U$. 
We then build the correlation matrix and use the $\chi^2$ implementation procedure
described in~\cite{Beniwal:2018hyi}. 
%
The region of the $(\sin \theta, m_s)$ plane excluded at the 2 $\sigma$ level is shaded blue in the right panel of Fig.~\ref{ChiSquare_Higgs}.\footnote{Our analysis does not include recent experimental measurements of $M_W$, which are discussed in~\cite{Bagnaschi:2022whn}.}

\subsection{LEP and LHC searches for BSM scalars}

The Higgs searches at LEP yield constraints on the singlet-doublet mixing angle $\theta$ for light singlet masses, $m_s \lesssim 100$ GeV (see e.g.~\cite{Robens:2015gla}), depicted in the right panel of Fig.~\ref{ChiSquare_Higgs} in the ($\sin \theta$, $m_s$) plane. Fig.~\ref{ChiSquare_Higgs} highlights that the LEP Higgs bounds do not yield relevant constraints on the parameter points of our scan, which are shown in black in the right panel of Fig.~\ref{ChiSquare_Higgs}.

In addition, BSM scalar searches at the LHC in $W W$, $Z Z$ and $h h$ decay channels constrain the properties of the singlet-like scalar state
$s$ for $m_s > m_h$. For $m_s > 200$ GeV, the strongest such limits have been 
obtained by ATLAS in the $Z Z \to 4\ell$ and $ Z Z \to 2\ell \, 2\nu$ final states~\cite{ATLAS:2020tlo} with $\sqrt{s} = 13$ TeV LHC data and 139 fb$^{-1}$ of integrated luminosity.
In order to interpret these bounds in our setup, we note that the effective
operator \eqref{eq:dim_5_Operator} impacts both gluon-gluon fusion production,
$g g \to s$, and the partial decay width of the singlet-like scalar into top
quarks $s \to t \bar{t}$, when phase space is available.
The strength of the interaction between the scalar $s$ and the top quark 
relative to that of the Higgs-top coupling in the SM is given by
\begin{equation}
\left|g_{stt} \right| =  \frac{1}{\sqrt{1 +  \frac{u^2}{\Lambda^2_{CP}}}} \sqrt{s_{\theta}^2 + \left( \frac{u\, s_{\theta}}{\Lambda_{CP}}  +  \frac{v\, c_{\theta}}{\Lambda_{CP}}\right)^2} 
\, ,
\end{equation}
where $s_{\theta} \equiv \sin\, \theta$ and $c_{\theta} \equiv \cos\, \theta$.
We note that $\left|g_{stt} \right| \to (v/\Lambda_{CP}) \times 1/\sqrt{1 +
(u/\Lambda_{CP})^2}$ in the limit $\sin\, \theta \to 0$, so the singlet production
cross section does not vanish in this limit, because of the
operator~\eqref{eq:dim_5_Operator} that is postulated for baryogenesis. 
Therefore LHC searches may be sensitive to very small singlet-doublet mixing,
if the value of $\Lambda_{CP}$ is close to the TeV scale. 
At the same time, we stress that if the $s \to t \bar{t}$ decay is kinematically accessible\footnote{We note that for very small mixing angles $\sin\, \theta < 0.01 - 0.001$ the three-body decay of $s$ via an off-shell (anti)top (open for $m_s \gtrsim m_t + m_W + m_b$) can still dominate over the $W W$ and $Z Z$ decays, whose partial widths are suppressed by $\sin^2\theta$ w.r.t. their SM-like values.} it becomes the dominant scalar branching fraction in the $\sin\, \theta \ll 1$ limit, suppressing the signal in $s \to Z Z$ searches. 
In such a case, the bounds from $p p \to s \to ZZ$ searches at the LHC become independent of the values of $u$ and $\Lambda_{CP}$, since the dependence on $\left|g_{stt} \right|$ approximately cancels in the limit of $\sin\, \theta \ll 1$ between $\sigma_{p p \to s} \propto \left|g_{stt} \right|^2$ and $\mathrm{BR}_{s \to Z Z} \propto \left|g_{stt} \right|^{-2}$. 
For scalar masses $m_s < 200$ GeV, LHC search results for $s \to WW,\,ZZ$ %
with $\sqrt{s} = 13$ TeV data are available from CMS~\cite{CMS:2018amk}, with a smaller (35.9 fb$^{-1}$) integrated luminosity.\footnote{LHC $s \to WW,\,ZZ$ searches with $\sqrt{s} = 7$ and $8$ TeV data~\cite{CMS:2013zmy,ATLAS:2015pre} yield much weaker constraints and do not affect the parameter points from our baryogenesis scan.} However, these still provide strong constraints in the mass region $130$ GeV $< m_s < 200$ GeV, where the ATLAS search~\cite{ATLAS:2020tlo} mentioned above does not apply.

The 2-$\sigma$ limits from BSM scalar searches at the LHC in $W W$ and $Z Z$ decay channels in the 
($\sin\, \theta$, $m_s$) plane are shown in the right panel of Fig.~\ref{ChiSquare_Higgs}, with the points
in our scan that are excluded by these LHC searches coloured red.
We note that $s$ may become long-lived in the limit of very small singlet-doublet mixing when the $s$ decay into top quarks (including the three-body decay $s \to t \,W \,b$) is kinematically forbidden, corresponding to $m_s \lesssim 260$ GeV, in which case the above $s \to ZZ$ limits from prompt LHC searches are evaded. 
%
Considering that the singlet scalar is typically produced with 3-momentum $\left| \vec{p}_s\right| \lesssim m_s$ (in the ATLAS/CMS detector frame) and the decay of $s$ ceases to be prompt for a decay-length $c\tau_s \gtrsim \mathcal{O}(\rm{mm})$, the ATLAS bounds from~\cite{ATLAS:2020tlo} and CMS bounds from~\cite{CMS:2018amk} are evaded for $\sin \theta \lesssim 10^{-6} - 10^{-7}$ (with the specific value decreasing as $m_s$ increases), as we depict in the right panel of Fig.~\ref{ChiSquare_Higgs}.

Finally, regarding future LHC searches, we stress that an increase in the sensitivity of searches for new scalars with $m_s < 200$ GeV in di-boson final states
(at present only the aforementioned CMS search~\cite{CMS:2018amk} with $\sqrt{s} = 13$ TeV LHC data exists)
could have the capability to explore fully the viable baryogenesis parameter space that remains unconstrained in the model. In particular, an estimate of the HL-LHC sensitivity in this mass region via a naive rescaling of the current CMS limits from~\cite{CMS:2018amk} by the square-root of the ratio of integrated luminosities 
indicates that mixing angles down to the long-lived singlet scalar case $\sin \theta \sim 10^{-6}$ could be probed in this mass range.


\subsection{The electron electric dipole moment}

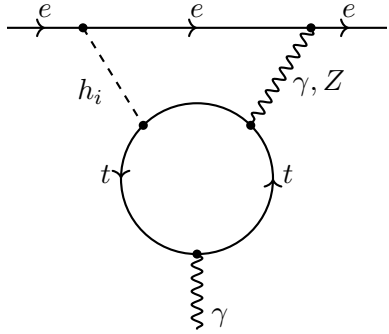
\begin{figure}[h]
\begin{center}
\begin{tikzpicture}[thick,scale=1.0]
\draw[particle] (-0.5,0) -- node[black,above,sloped,yshift=-0.0cm,xshift=-0.0cm] {$e$} (0.5,0);
\fill[black] (0.5,0.0) circle (0.06cm);
\draw[particle] (0.5,0) -- node[black,above,sloped,yshift=-0.0cm,xshift=0.0cm] {$e$} (3.5,0);
\draw[particle] (3.5,0) -- node[black,above,sloped,yshift=-0.0cm,xshift=0.0cm] {$e$} (4.5,0);
\draw[dashed] (0.5,0) -- node[black,above,yshift=-0.5cm,xshift=-0.3cm] {$h_i$} (1.293,-1.293);
\fill[black] (1.293,-1.293) circle (0.06cm);
\draw[decorate,decoration={snake,amplitude=2pt,segment length=5pt}] (3.5,0) -- node[black,above,yshift=-0.4cm,xshift=0.5cm] {$\gamma,Z$} (2.707,-1.293);
\fill[black] (2.707,-1.293) circle (0.06cm);
\draw[decorate,decoration={snake,amplitude=2pt,segment length=5pt}] (2,-3) -- node[black,above,yshift=-0.6cm,xshift=0.3cm] {$\gamma$} (2,-4);
\fill[black] (2,-3) circle (0.06cm);
\draw[particle]  (2,-1) node[black,above,sloped,yshift=-1.2cm,xshift=-1.2cm] {$t$}  arc (90:270:1cm) ;
\draw[particle]  (2,-3) node[black,above,sloped,yshift=0.8cm,xshift=1.2cm] {$t$}  arc (-90:90:1cm) ;
\fill[black] (3.5,0) circle (0.06cm);
\end{tikzpicture}
\end{center}
\caption{\it The $2$-loop Bar-Zee diagram that contributes to the
electron electric dipole moment~\cite{Barr:1990vd}. Here $h_i$ 
denotes the mass eigenstates $h$ and $s$.}
\label{eEDM_diagram}
\end{figure}

The effect of the dim-$5$ operator on the electron electric dipole 
moment (eEDM) is given by the $2$-loop Bar-Zee
diagram~\cite{Barr:1990vd} with a top quark loop,
see Fig.~\ref{eEDM_diagram}, whose contribution is given by the 
following formula \cite{Keus:2017ioh}:
\begin{equation}
    d_e^{2-\text{loop}} =  \frac{e}{3\pi^2}\frac{\alpha G_F v}{\sqrt{2}\pi m_t}m_e \left(  \frac{v}{\sqrt{2}\Lambda_{\text{CP}}}\right)\sin{\theta}\cos{\theta} \left[ -g\left( \frac{m_t^2}{m_h^2}\right) + g\left( \frac{m_t^2}{m_s^2}\right) \right] \, ,
    \label{eqn:eEDM_formula}
\end{equation}
with the $1$-loop integral
\begin{equation}
    g(z) =\frac{z}{2}\int_0^1 dx \frac{1}{x(1-x)-z}\log{\left[  \frac{x(1-x)}{z}\right]} \, .
\end{equation}
We see in the above expression that the prediction for the eEDM 
depends only on $\theta$, $m_s$ and the cutoff scale $\Lambda_{CP}$, and in particular it vanishes in the limit of vanishing mixing angle $\theta \to 0$ or scalar mass degeneracy $m_s \to m_h$.

We use the following numerical values for the physical constants \cite{ParticleDataGroup:2020ssz} in the formula for the eEDM:
\begin{align}
    e&\equiv g' =0.34 \, , \nonumber \\
    \alpha &= 1/137 \, , \nonumber \\
     G_F&\equiv\frac{1}{\sqrt{2}v^2} = 1.166 \times 10^{-5} (\text{GeV})^{-2} \, ,  \nonumber \\
      m_e&=0.5 \ \text{MeV} \, , \nonumber \\
       m_t&=172.9 \ \text{GeV} \, .
\end{align}
The best current experimental upper bound on the eEDM, obtained 
by the ACME collaboration~\cite{ACME:2018yjb}, is
\begin{equation}
    |d_e|<1.1 \times 10^{-29} e \ \text{cm} = 1.89 \times 10^{-16} \ \text{GeV}^{-1} \, .
\end{equation}

\subsection{The baryon asymmetry}\label{sec:BAU}



Although the above formula for the field-dependent top quark mass might
seem like a minor modification of the scalar potential, it can
nevertheless have a significant numerical impact on the thermodynamic parameters of
the transition, and must be taken into account. These effects have not
been considered in previous studies. Instead, it has been customary to
neglect altogether the impact that the 
{mixing with the scalar}
in 
Eqn.~\eqref{eq:dim_5_Operator} has on the phase transition and to solve
for the value of $\Lambda_{CP}$ that yields the required BAU.
We note in addition that the dependence of the top quark mass on the singlet 
profile induces a friction term, $F_{\text{friction}} \propto d m_t/ds$,
in the singlet equation of motion. However, we have verified that the
numerical impact of this term is irrelevant.

Since the above modifications lead to a different scalar potential from
that studied in the preceding Sections, we have performed an independent
scan of the parameter space with the above modified effective potential,
treating $\Lambda_{CP}$ as one of its parameters, using the eEDM constraint as a filter
{and incorporating all the phenomenological constraints.}
We find that the new
effective potential always develops a global minimum with $(h,s)=(v,u)$
that is distinct from that at the
electroweak scale, but that its location in
field space is several orders of magnitude beyond the Planck scale. Hence we do not consider this prospective destabilization to be a serious
concern. 

\begin{figure}[!h]
\centering
\includegraphics[scale=.35]{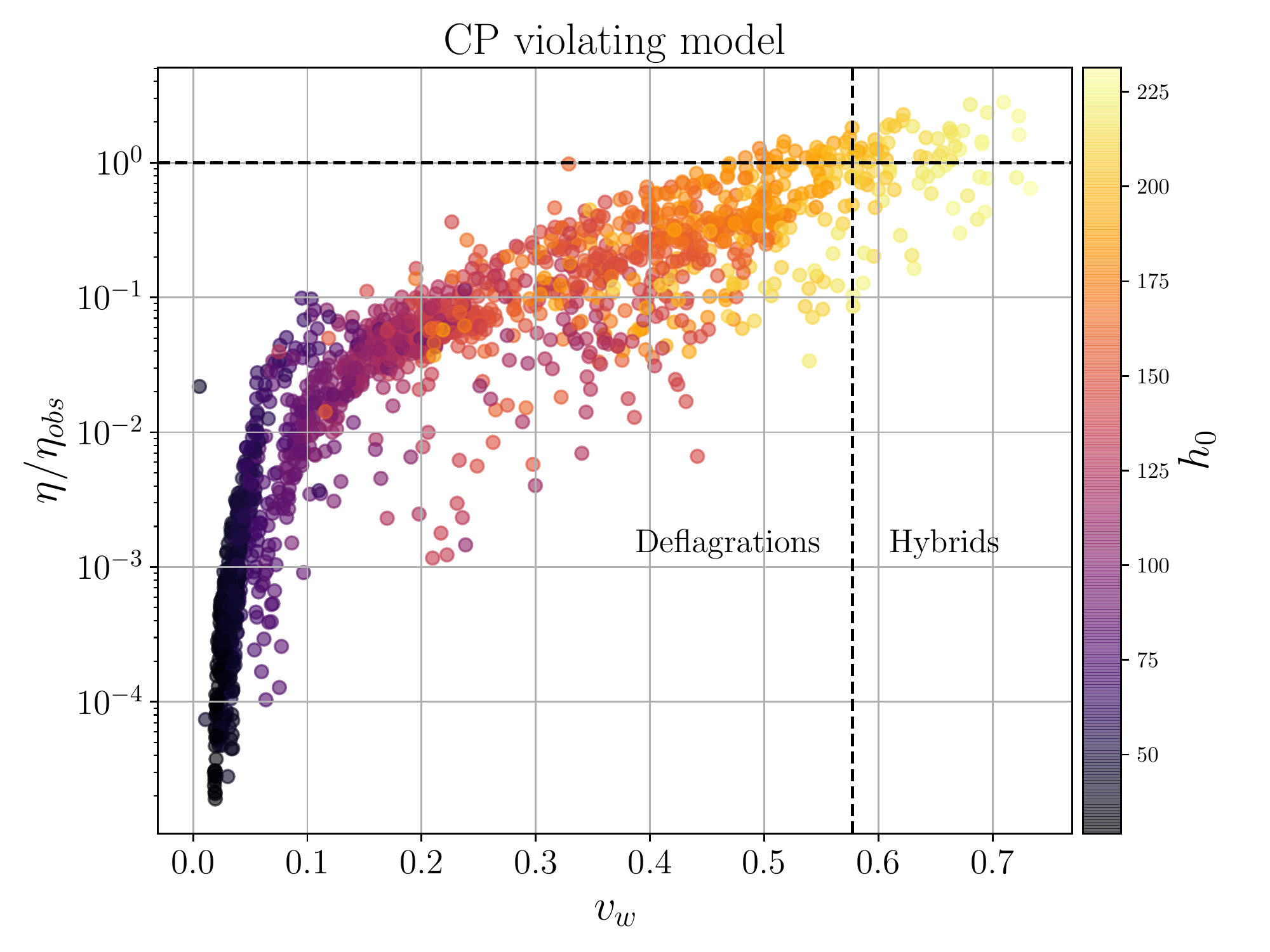} 
\includegraphics[scale=.35]{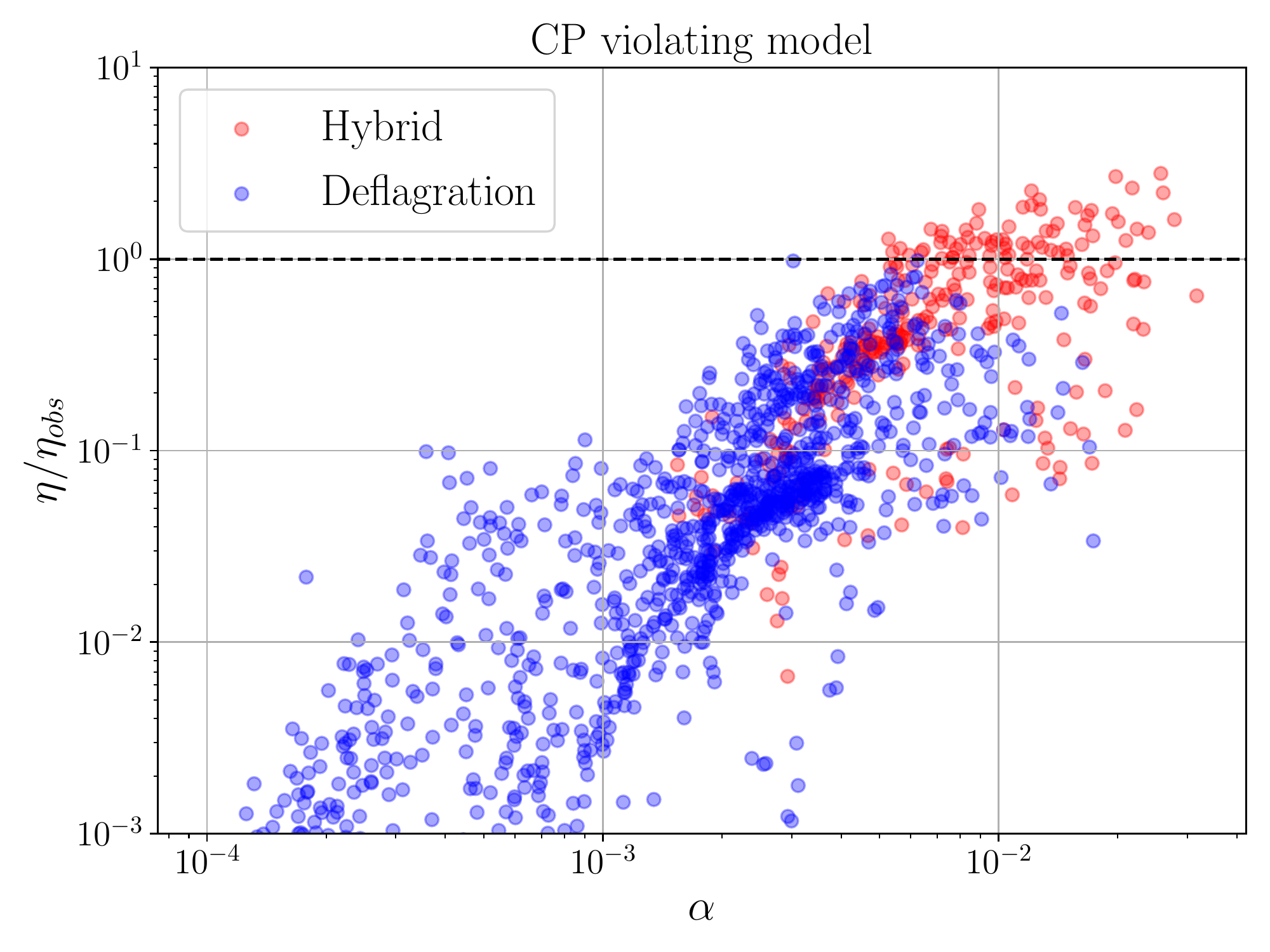}
\caption{\it Results for the baryon asymmetry of the Universe 
(BAU) normalized to its observed value as a function of the 
wall velocity (left panel) and the strength $\alpha_0$ of the phase transition (right panel). The coloured side-bar 
alongside the left panel indicates the value of the Higgs bubble profile amplitude $h_0$.}
\label{BAU_vw}
\end{figure}

 To compute the baryon asymmetry of the Universe (BAU) we use the improved 
fluid equations of Ref.~\cite{Cline:2020jre}, which are well 
behaved for any value of the wall velocity. 
We have scanned over parameter values uniformly distributed in the following
ranges:
\begin{align}
m_s    &\ \in \ [1,1000 ]\ \text{GeV} \, , \nonumber \\
\theta &\ \in \ [-0.1,0.1] \, ,  \nonumber \\
\Lambda_{CP} &\ \in \ [v,1000 ]\ \text{GeV} \, , \nonumber \\
u &\ \in \ [-1000,1000 ]\ \text{GeV} \, , \nonumber \\
\mu_3 &\ \in \ [-1000,1000 ]\ \text{GeV} \, , \nonumber \\
\mu_{hs} &\ \in \ [-1000,1000 ]\ \text{GeV} \, .
\end{align}
{As discussed above, we have checked that all the points are consistent with the phenomenological constraints, and that} $\Lambda_{CP}$ is the highest
mass scale, so that the effective field theory is always consistent.
The results for the full computation are presented in Fig.~\ref{BAU_vw}
as functions of the wall velocity $v_w$ and the strength parameter $\alpha_0$ 
(coloured in the left panel according to the value of $h_0$), 
which are the parameters with the most significant impact. 
As indicated, the BAU $\eta$ is positively correlated with the 
wall velocity and with the Higgs bubble profile amplitude, in agreement with the findings in \cite{Lewicki:2021pgr}. 
We see that $\eta$ can attain its observed value today most easily when the wall expansion is hybrid. 

\begin{figure}[!h]
\centering
\includegraphics[scale=.4]{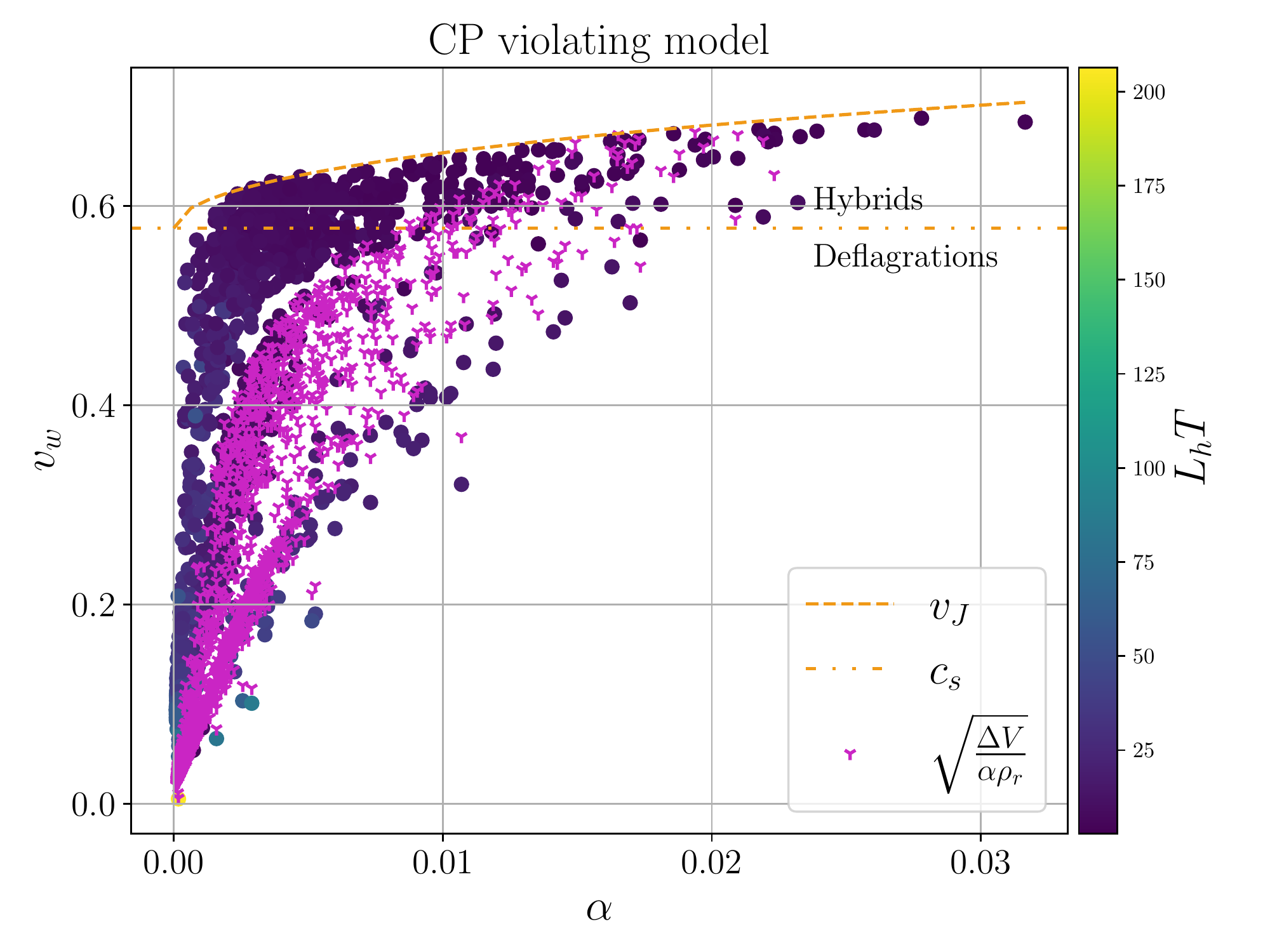} 
\caption{\it The wall velocity as a function of the strength 
parameter $\alpha$ for the CP-violating model. The lines and
the colour coding of the points are the
same as in Fig.~\ref{singlet_bubble}.}
\label{vw_CP}
\end{figure}

\begin{figure}[!h]
\centering
\includegraphics[scale=.6]{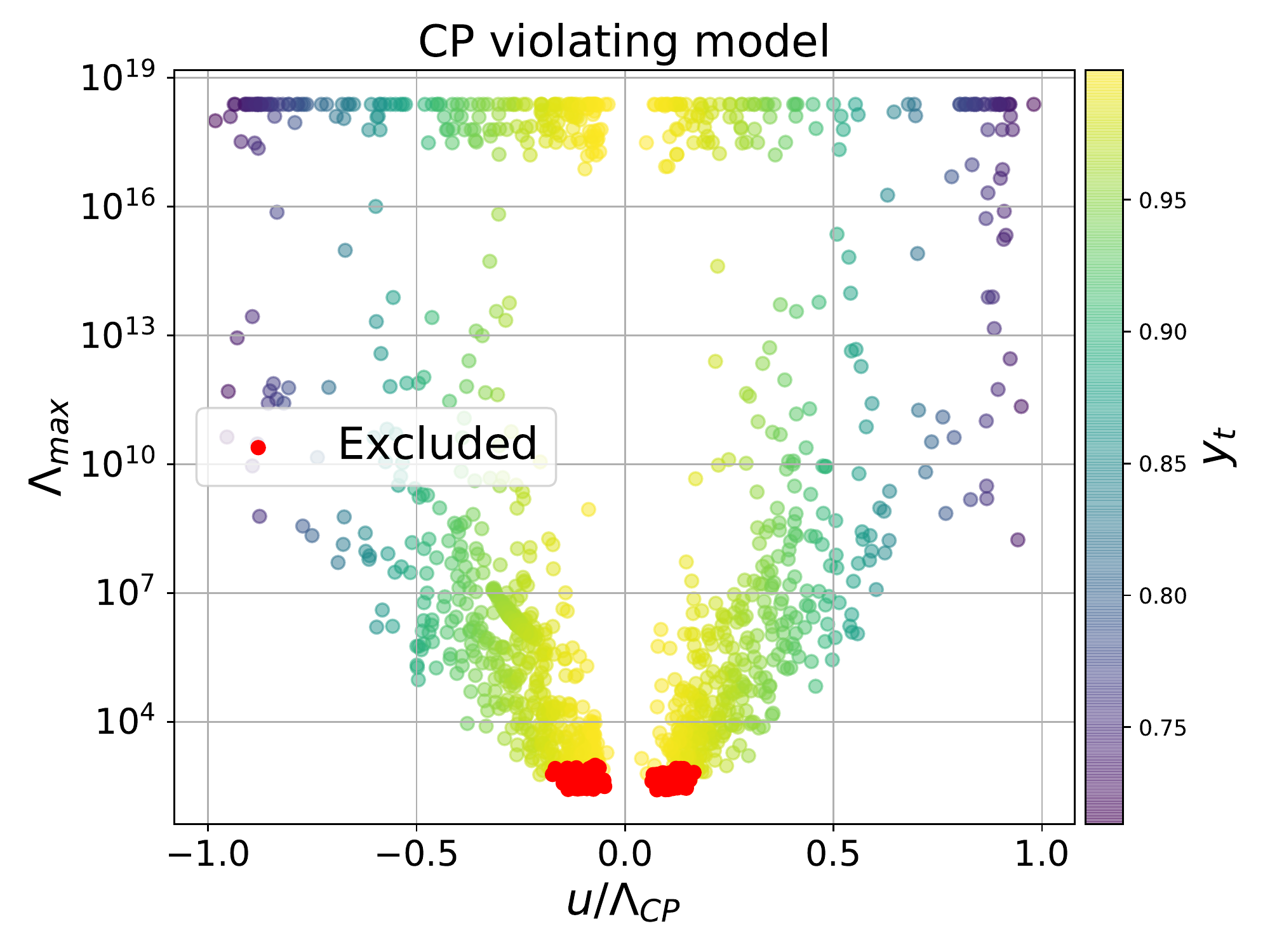} 
\caption{\it Maximum renormalization scale where positivity and perturbativity are satified as a function of $u/\Lambda_{CP}$. 
The coloured side-bar shows the value of the top Yukawa coupling.}
\label{RGE_plot}
\end{figure}

\begin{figure}[!h]
\centering
\includegraphics[scale=.65]{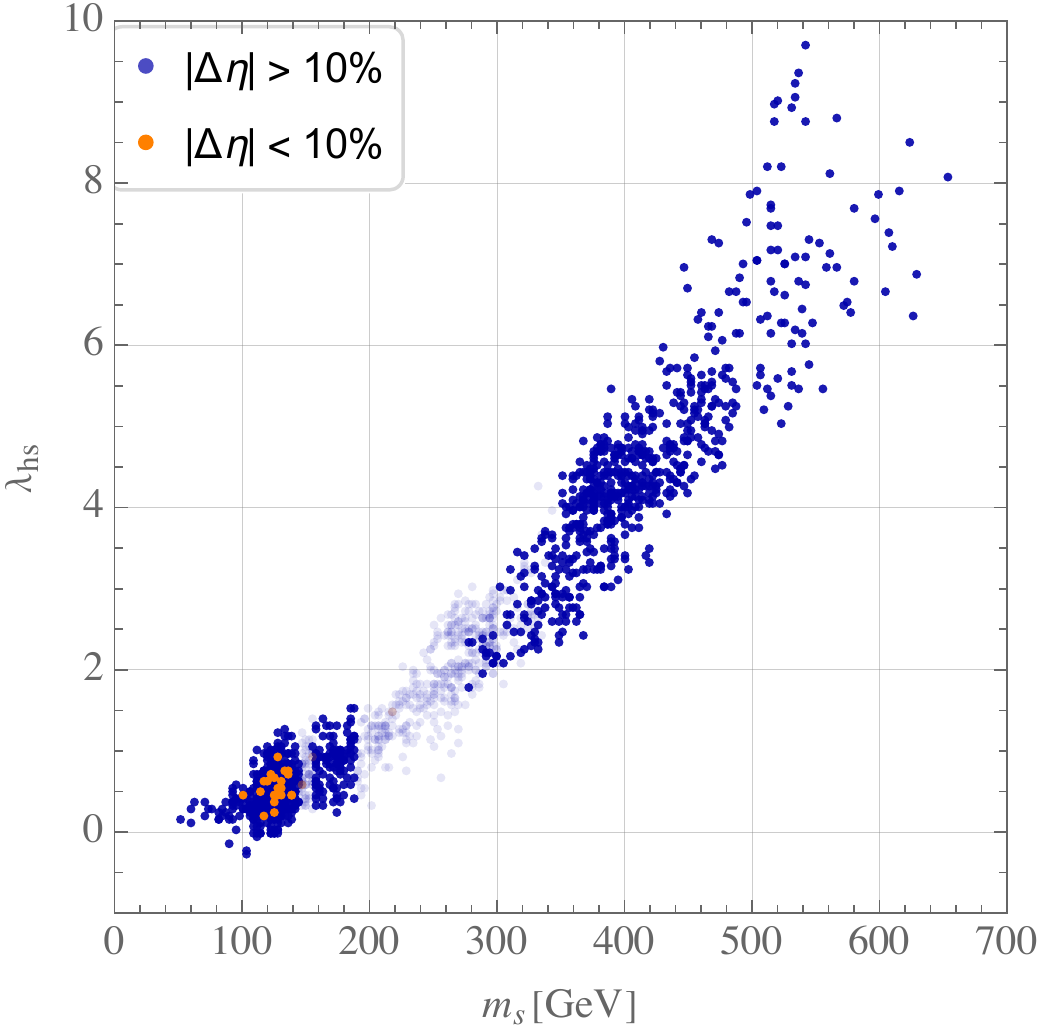}
\includegraphics[scale=.65]{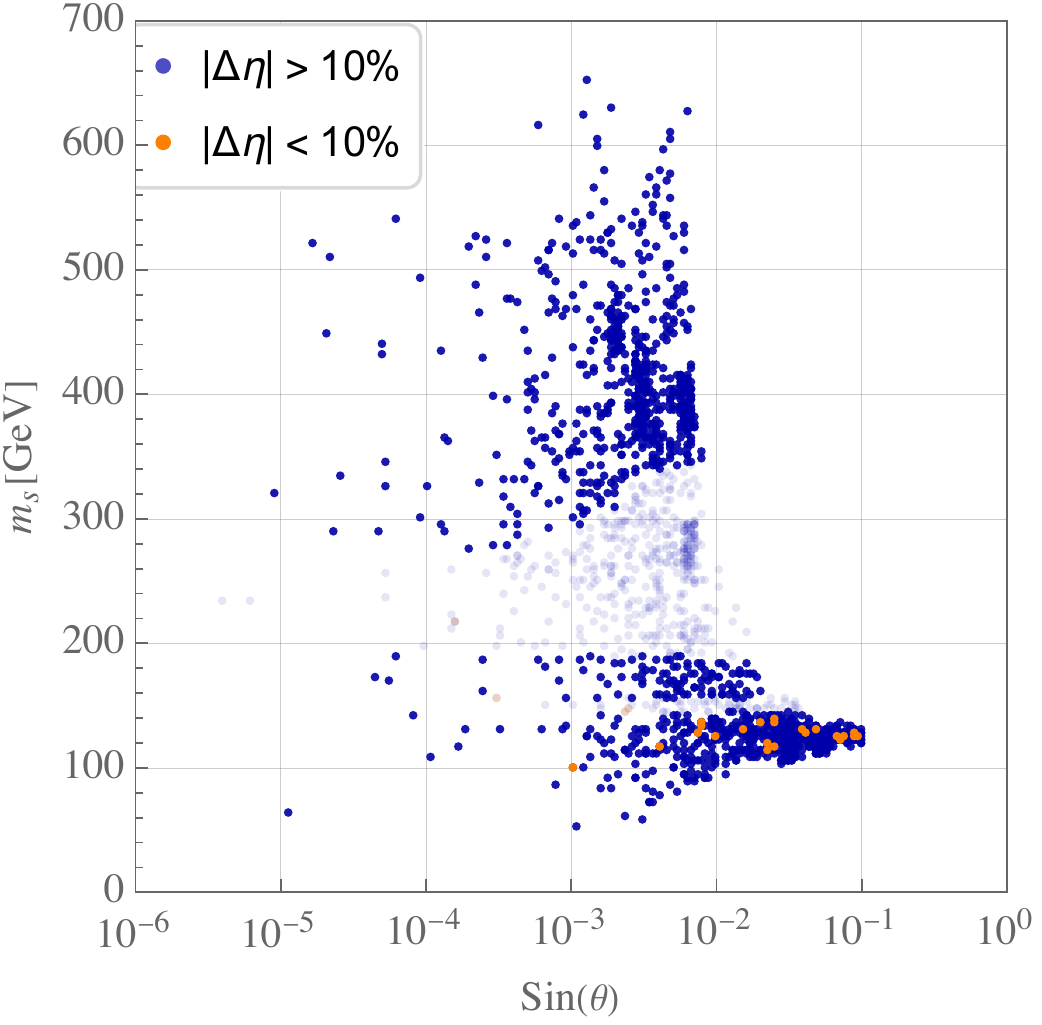}
\caption{\it Allowed parameter space of the CP-violating singlet model in the $(m_s, \lambda_{hs})$ plane (left panel) and the $(\sin \theta, m_s)$ plane (right panel). The orange points predict a baryon asymmetry that lies within $10\%$ deviation
of the cosmological value, and hence delimit the preferred region.
}
\label{CP_paramspace}
\end{figure}

In analogy with our results for the bubble wall properties shown
in the previous Section, we show in Fig.~\ref{vw_CP} the results
for the CP-violating model. We see that the range 
of the strength parameter $\alpha$ is significantly smaller than 
in the generic model without CP violation. As a consequence, the 
analytic formula yields underestimates for all the points and,
as we saw in Fig.~\ref{relative_error}, the relative error is 
significant.
Because of the low values of $\alpha$, the
GW signal is too small to be measured in
planned experiments.
In summary: when the properties of the wall are calculable the GW signals are not strong enough to be measurable 
in forthcoming experiments, and the GW spectra in the models with and without CP violation look very similar.

It is well known that in the SM the perturbative effective potential develops an instability at a scale
$\sim 10^{12}$~GeV induced by the negative renormalization of the Higgs quartic coupling that is induced by the top quark
Yukawa coupling. Therefore the rescaling of the top Yukawa rescaling raises a natural question: 
What is the impact of this rescaling on the stability and perturbativity of the potential at higher energy scales?
A dedicated answer lies beyond the scope of the present paper, but we have investigated this question 
using the 1-loop renormalization group equations, as described in an Appendix. As seen in Fig. \ref{RGE_plot}, we have found
that only a modest number of points (coloured red) are ruled out because instability sets in below $\Lambda_{CP}$,
that there are many points which are no more unstable than the SM, and that there are a sizeable fraction of points
that are stable and theoretically consistent up to the Planck scale. 

Finally, we display in the left panel of Fig.~\ref{CP_paramspace} the parameter space in the plane of singlet mass $m_s$ and Higgs portal coupling $\lambda_{hs}$
that is consistent with all experimental data and with a non-zero $\eta \neq0$, and in the right panel we show the allowed parameter space in the $(\sin \theta, m_s)$ plane.
The sparseness of the sample for $190$ GeV $\lesssim m_s \lesssim 300$~GeV and in the small region around $m_s \sim 150$~GeV is due to the
impact of the LHC constraints. The orange points give $\eta$ within $10 \%$ of 
the observed value today and thus constitute the most successful predictions found
in the present study. The relatively small mass of the $s$
boson for such scenarios, $m_s \lesssim 150$~GeV, is within the
kinematic reach of the LHC for singlet-like scalar production in association with a $\bar t t$ pair, and will be also probed by future LHC searches in di-boson final states. 
We also note that increasing the precision of the eEDM experiment by an order of magnitude (as expected by the ACME Collaboration in the future~\cite{ACME:2018yjb,Wu:2019jxj}) would be sensitive to around 75\% of our sample of  points that yield the right baryon abundance. In contrast, as already mentioned, the GW signal is too small to be observable. 

\section{Conclusions}
\label{Sec:conx}

We have revisited in this paper the extension of the SM with one additional gauge singlet scalar field,
allowing the $\mathbb{Z}_2$ symmetry of the potential to be broken. We have considered both the case 
where the symmetry is broken spontaneously and the more general case in which
the potential is extended with extra terms and the symmetry breaking is explicit.
We have investigated details of the electroweak phase transition
in the different cases, with the aim of estimating the possible magnitude of the gravitational wave signal,
assessing the possibility of realising electroweak baryogenesis, and investigating whether they can coexist.

In the case with spontaneous breaking we find that after inclusion of the full one-loop potential the transitions are always weak, i.e., $v/T \lesssim 1$, which gives little hope for electroweak baryogenesis. The case with explicit symmetry breaking 
is more promising, and may give rise rise to a transition strong enough for baryon production. We investigate the bubble wall properties, using the semiclassical fluid equations to calculate the final velocity and shape of the wall. This allows us to compute accurately the baryon yield, for which we use a dimension-five operator that couples the singlet scalar to the SM top quarks, yielding a new source of CP violation.
Despite stringent constraints from the experimental limits on 
the electron dipole moment and from direct searches for singlet-like scalars at the LHC, we find that a significant part of the allowed parameter space produces a baryon yield close to the observed value. Such regions of parameter space are within reach of upcoming LHC searches, and could also be probed in the future by an $\mathcal{O}(10)$ increase in the sensitivity of electron electric dipole moment experiments.

However, just as in other simple SM extensions~\cite{Cline:2021iff,Lewicki:2021pgr}, we find that all the transitions capable of producing an appreciable baryon yield result in GW signals that are too weak to be observed in upcoming experiments. 
The reason is that, already for weak transitions with $\alpha\approx 0.05$, the vacuum pressure becomes sufficiently large 
that the wall is not stopped in our semiclassical fluid picture. To be more precise, there is a distinction between weak transitions 
for which the hydrodynamical solution is such that the plasma is heated in front of the wall and stronger transitions 
that predict a so-called detonation solution in which the plasma outside the wall has the same temperature as before the transition. 
If the transition is strong enough that the heated plasma shell outside the wall disappears the friction drops substantially 
and there is no solution  up to very large wall velocities where the semi-classical fluid approximation breaks down. This means that for such strong transitions the wall will reach a velocity close to the speed of light 
and the hope of realising electroweak baryogenesis fades. Because the division between the two cases is at such small transition strength,
we find that only transitions with relativistic walls predict observable gravitational wave signals.   

As a by-product we provide an analytical approximation for the wall velocity which does not require any additional
computation beyond those of the standard thermodynamics parameters used when approximating the GW signal. 
We have compared this approximation with our detailed numerical results, and find that it gives results 
that are typically accurate with within a few percent. We also find it to be accurate for separating the solutions 
between those with slow walls that are appropriate for baryogenesis and those with relativistic walls that give strong GW signals.   

\section*{Acknowledgements}
We would like to thank Bogumi\l a \'Swie\.zewska for comments on the manuscript.
The work of JE was supported in part by United Kingdom STFC Grants ST/P000258/1 and ST/T000759/1, and in part by the Estonian Research Council via a Mobilitas Pluss grant.
The project is co-financed by the Polish National Agency for Academic Exchange within Polish Returns Programme under agreement PPN/PPO/2020/1/00013/U/00001 and the Polish National Science Center grant 2018/31/D/ST2/02048.
The work of J.M.N. is supported by the Ram\'on y Cajal Fellowship contract RYC-2017-22986, and by grant PGC2018-096646-A-I00 from the Spanish Proyectos de I+D de Generaci\'on de Conocimiento.
J.M.N. also acknowledges support from the European Union's Horizon 2020 research and innovation programme under the Marie Sklodowska-Curie grant agreement 860881 (ITN HIDDeN), as well as from  
the grant IFT Centro de Excelencia Severo Ochoa
CEX2020-001007-S funded by MCIN/AEI/10.13039/501100011033. 

\appendix

\section{One-Loop Analysis of Positivity and Perturbativity}
The beta functions in the most generic scalar singlet extension without $\mathbb{Z}_2$ symmetry have been studied in Ref. \cite{Ghorbani:2021rgs} in the context of vacuum stability and positivity. In this paper we are only concerned about positivity and perturbativity of the quartic couplings, thus the relevant RGE system at 1-loop is given by 
\begin{align}
16 \pi^2 \beta_{g_1} & = \frac{41}{6} g_1^3,\\
16 \pi^2 \beta_{g_2} & =  -\frac{19}{6} g_2^3, \\
16 \pi^2 \beta_{g_3} & =  -7 g_3^3, \\
16 \pi^2 \beta_{y_t} & =  \frac{9}{2} y_t^3 - 8g_3^2 y_t - \frac{9}{4}g_2^2 y_t - \frac{17}{12}g_1^2 y_t, \\
16 \pi^2 \beta_{\lambda} & =  24 \lambda^2 + \frac{\lambda_{hs}^2}{2} - 6 y_t^4 + 12 y_t^2\lambda  - 9 \lambda g_2^2 - 3 \lambda g_1^2 + \frac{3}{8}g_1^4 + \frac{3}{4}g_2^2 g_1^2 + \frac{9}{8}g_2^4, \\
16 \pi^2 \beta_{\lambda_{hs}} & = 4 \lambda_{hs}^2 + 6\lambda_s \lambda_{hs} - \frac{9}{2}\lambda_{hs}g_2^2 -\frac{3}{2}\lambda_{hs} g_1^2 + 6 \lambda_{hs} y_t^2 + 12\lambda_{hs}\lambda , \\
16 \pi^2 \beta_{\lambda_s} & = 2\lambda_{hs}  + 18 \lambda_s^2,
\end{align}
with  
\begin{equation}
\beta_g = \frac{d g}{dt}, \quad t =\log{\mu},
\end{equation}  
where $g_1$, $g_2$, $g_3$ are the $U(1)_Y$, $SU(2)_L$ and $SU(3)_c$ gauge couplings. 
Following \cite{Casas:1994us} we impose as initial conditions
\begin{align}
g_1(M_Z) &= 0.344, \\
g_2(M_Z) &= 0.64, \\
g_3(M_Z) &= 1.22.
\end{align}
We solve the RGE evolution equations from the electroweak scale $\mu=246$ GeV up to a maximum scale $\mu=\Lambda_{max}$ at which the quartic couplings fulfill the positivity and perturbativity conditions. We consider the point to be ruled out if $\Lambda_{max}<\Lambda_{CP}$.

\bibliographystyle{JHEP}
\bibliography{Baryogen.bib}  
  
\end{document}

\begin{figure}[hbt!]
\centering
\subfigure{
\includegraphics[width=0.45\textwidth]{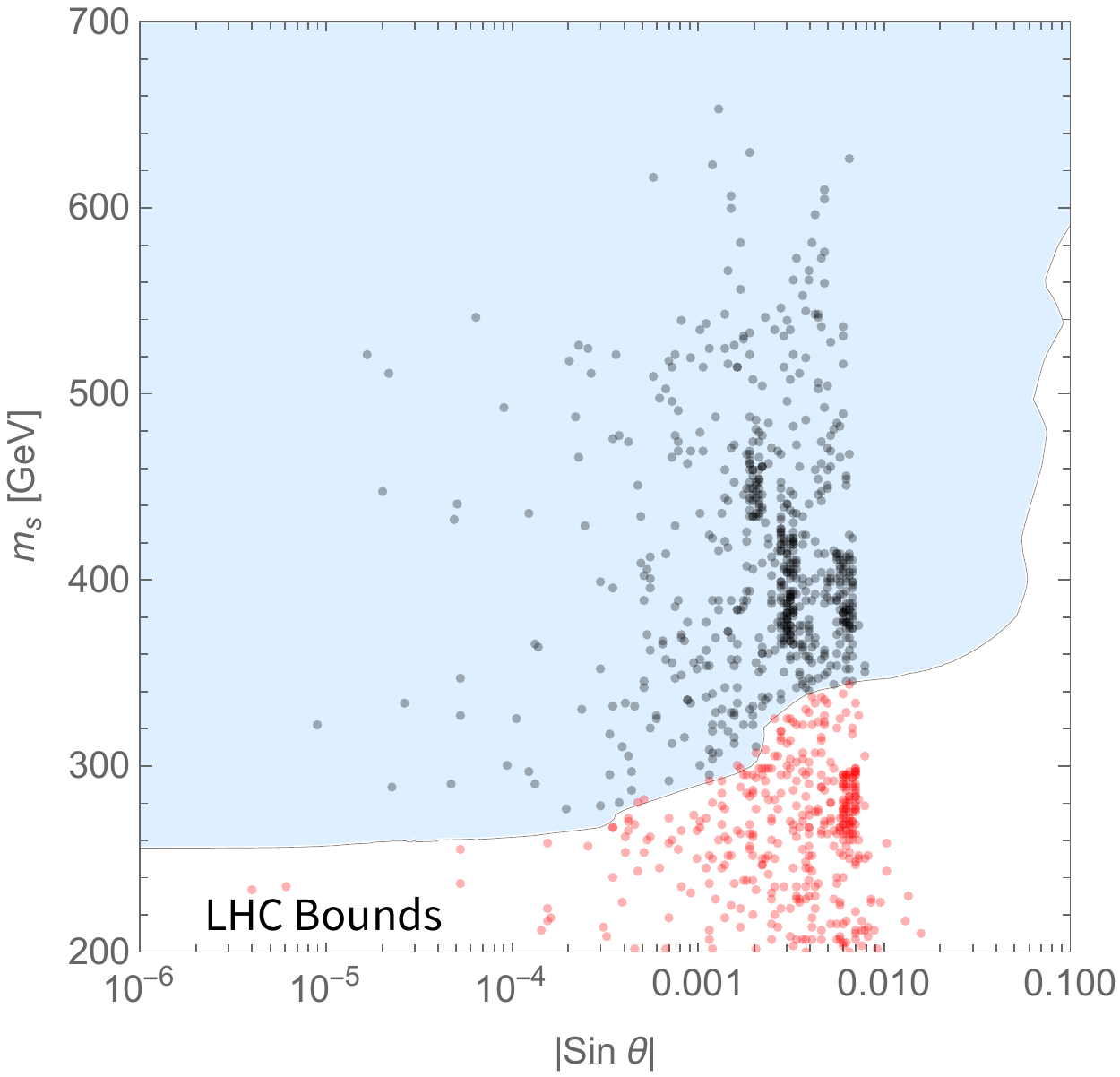}
}
\caption{\it 2-$\sigma$ allowed regions in the $(\sin \theta, m_s)$ plane for $u= 100$ GeV and $\Lambda_{CP} = 1000$ GeV from the ATLAS search for ZZ~\cite{ATLAS:2020tlo}.} 

\label{LHC_singlet_ATLAS}
\end{figure}

\red{still need to explain the plot!}